\documentclass[aps,prd,reprint,amsmath,noeprint]{revtex4-2}
\usepackage{dcolumn}
\usepackage{orcidlink} 
\hypersetup{
    colorlinks=true,    
    linkcolor=[RGB]{45, 47, 145}, 
    citecolor=[RGB]{45, 47, 145}, 
    filecolor=[RGB]{45, 47, 145}, 
    urlcolor=[RGB]{45, 47, 145}   
}

\begin{document}

\title{Revisiting $B\to K^*\pi$ and $B\to K\rho$ decays using the modified perturbative QCD approach}
\author{Ru-Xuan Wang\,\orcidlink{0000-0002-6394-1225}} \email{wangrx@mail.nankai.edu.cn} 
\author{Mao-Zhi Yang\,\orcidlink{0000-0002-5052-713X}} \email{yangmz@nankai.edu.cn} 
\affiliation{School of Physics, Nankai University, Tianjin 300071, People's Republic of China}
\date{\today}

\begin{abstract} 
We revisit $B\to K^*\pi$ and $B\to K\rho$ decay processes using the modified perturbative QCD approach, which is developed based on the perturbative QCD approach with a few improvements. 
A critical infrared cutoff scale $\mu_c$ is introduced to separate hard and soft contributions in the decay process. 
Hard contributions with scale $\mu$ larger than $\mu_c$ are calculated with perturbative QCD, and soft contributions with $\mu<\mu_c$ are described in terms of soft transition and production form factors. 
In addition, color-octet contributions are considered phenomenologically. The branching fractions and $CP$ violations of  $B\to K^*\pi$, $K\rho$ decays are calculated. 
With appropriate parameter inputs, the theoretical results can be in good agreement with experimental data. 
\end{abstract}

\maketitle

\section{Introduction \label{sec:intro}} 
$B$ factories and LHCb experiment have collected large amount of precise data on $B$ meson decays \cite{ParticleDataGroup:2024cfk}, 
which provide a strong basis for investigating the rich physics involved in strong and weak interactions in $B$ decays. 
In theory, treating strong interaction effect in QCD is a challenging task. Factorization theorem is usually used to separate physics of different scales in QCD. 
The amplitude of an exclusive process can be expressed as a convolution of hadron wave functions and hard transition function of subprocess at short-distance in QCD \cite{Lepage:1980fj}.  
And it was proven a long time ago that all nonperturbative contributions can be absorbed into the definition of the $B$ meson and light meson distribution amplitudes in $B$ decays \cite{Li:1994cka,Li:1995jr,Li:1994zm,Li:1994iu,Chang:1996dw,Yeh:1997rq,Nagashima:2002ia}. 
This is the perturbative QCD (PQCD) approach.

Recently, to resolve the serious tension between experimental data and theoretical predictions in $B\to \pi\pi$ and $K\pi$ decays, 
we introduced the critical infrared cutoff scale $\mu_c$ in the calculation of $B$ decay amplitude, 
where the contributions with the scale higher than $\mu_c$ are calculated in PQCD approach, 
while dynamics with scale $\mu<\mu_c$ is attributed to several soft transition and production form factors \cite{Lu:2021gwt,Lu:2022hbp,Wang:2022ihx}. 
In addition, soft color-octet contributions are introduced in the calculation of the decay amplitudes \cite{Lu:2022hbp,Wang:2022ihx}. 
This modified PQCD approach is extended to all $B\to PP$ decays, where $P$ denotes pseudoscalars \cite{Lu:2024jzn}. 
By using SU(3) flavor symmetry and its breaking effects, 
color-octet contributions can be uniformly considered and the branching ratios and $CP$ violations for all the $B\to PP$ decays can be in good agreement with experimental data. 

In this work, we revisit $B\to K^*\pi$ and $B\to K\rho$ decays with this modified PQCD approach. 
The branching ratios and $CP$ violations are calculated and compared with experimental data. 
With the introduction of the infrared cutoff scale $\mu_c$, the applicability of the perturbative calculation is improved. 
Compared with leading-order (LO) contributions in QCD, next-to-leading order (NLO) contributions only slightly change the branching ratios in general. 
Only for a few decay modes, the NLO contribution can affect the branching ratios as much as $20\%\sim 30\%$. 
The soft contribution, especially the color-octet contribution, is significant for explaining the experimental data. 
With appropriate parameter inputs, all the physical results can be in good agreement with experimental data.

The remaining part of the paper is organized as follows. 
In Sec.~\ref{sec:amp}, we give the perturbative calculation of the hard decay amplitude with PQCD approach. 
In Sec.~\ref{sec:soft}, we present the soft contributions from soft form factors and color-octet matrix elements. 
Sec.~\ref{sec:num} is devoted to numerical analysis and discussion, and Sec.~\ref{sec:sum} is a brief summary.

\section{Decay amplitudes at leading and next-to-leading order \label{sec:amp}}
\subsection{Factorization formalism \label{subsec:fac}}
The nonleptonic $B$ meson decay involves various energy scales, including the electroweak scale, $B$ meson scale, hard scattering scale, and QCD scale. 
Direct calculation for these decays proves to be quite challenging. Typically, factorization approach is utilized to study such multi-scale questions. 
Factorization means the physical amplitude is decomposed into several parts according to the energy scales of the interactions involved. 
For processes involving QCD, we can only calculate the perturbative part in the hard-scale (short-distance) region perturbatively due to the asymptotic freedom, 
while nonperturbative contribution must be calculated with nonperturbative methods or extracted from experimental data. 
In the factorization approach, the nonperturbative contribution is absorbed into the meson wave function, $\Phi$, which characterizes the hadronization of quark-antiquark pairs. 
This wave function can be extracted from the experimental data or nonperturbative methods and is not dependent on the specific decay process. 

The $B\to K^*\pi$ and $B\to K\rho$ decays are induced by charged weak current within the Standard Model. 
By integrating out the heave $W$ boson, this transition can be appropriately described by the four-quark operators $b\to sq\bar{q}$. 
In the center-of-mass frame of the $B$ meson, the momenta of the two outgoing mesons are large and have the opposite directions, 
since the $B$ meson is considerably heavier than the final-state mesons. To generate the large recoiling final-state meson, the light (spectator) quark within the $B$ meson, 
whose momentum is in QCD scale $\Lambda_{\text{QCD}}$ initially, must undergo a hard gluon exchange process with the quark in the weak four-quark operator. 
These decays are dominated by the hard process and can be effectively studied in the framework of PQCD approach \cite{Lu:2000em, Keum:2000wi}. 
Furthermore, the $k_T$ factorization is necessary to regulate the end-point singularity, where $k_T$ is the transverse momentum of the quark. 
The extra $k_T$ scale introduces the extra logarithms which spoil the perturbative expansion. 
The double logarithms $\alpha_s\ln^2k_T$ and $\alpha_s\ln^2x$ should be resummed into the Sudakov factor $s(P,b)$ \cite{Li:1992nu} and threshold resummation factor $S_t$ \cite{Li:2001ay}, respectively. 

Formally, the $B\to M_1M_2$ decay amplitude can be expressed by the convolution 
\begin{align} \label{eq:factorization}
	\mathcal{M} &= \int dk_0dk_1dk_2 \Phi^B(k_0,t) C(t) \nonumber \\
		&\quad \times  H(k_0,k_1,k_2,t) \Phi^{M_1}(k_1,t) \Phi^{M_2}(k_2,t) \nonumber \\
		&\quad \times \exp\left[-s(P,b)-2\int_{1/b}^t \dfrac{d\bar{\mu}}{\bar{\mu}}\gamma_\Phi(\alpha_s(\bar{\mu}))\right]. 
\end{align}
The Wilson coefficient $C(t)$ at the energy scale $t$, derived from the renormalization group method, contains the contributions from heavy particles. 
The hard kernel $H$, which can be calculated perturbatively, denotes the contributions from the hard scale region. 
The meson wave functions $\Phi^{B(M_1, M_2)}$ include the contributions down to the $\Lambda_{\text{QCD}}$ scale. And $\gamma_\Phi$ represents the anomalous dimension for the wave functions. 

\subsection{Effective Hamiltonian \label{subsec:heff}}
Given the heavy mass of the weak gauge boson $W$ compared with the $B$ meson, it's safe to integrate out the $W$ boson and derive the effective Hamiltonian for four-quark operators. 
For $b\to sq\bar{q}$ decays with $q\in\{u,d\}$, as encountered in $B\to K^*\pi$ and $B\to K\rho$ decays, the corresponding effective Hamiltonian is expressed as \cite{Buchalla:1995vs} 
\begin{align}\label{eq:heff}
	\mathcal{H}_{\mathrm{eff}} =&\dfrac{G_F}{\sqrt{2}}\biggl\{V_u\left[C_1(\mu)O_1^u(\mu) + C_2(\mu)O_2^u(\mu)\right] \nonumber \\
		&\quad -V_t\sum_{i=3}^{10}\left[C_i(\mu)O_i(\mu)+C_{8g}(\mu)O_{8g}(\mu)\right]\biggr\}, 
\end{align}
with $G_F$ being the Fermi constant, the Cabibbo-Kobayashi-Maskawa (CKM) matrix elements $V_u=V_{ub}V_{us}^*$ and $V_t=V_{tb}V_{ts}^*$, 
$C_i(\mu)$ the Wilson coefficients, $O_i(\mu)$ 
the four-quark operators, and $\mu$ the factorization scale. 
The local four-quark operators are defined as follows: 
{
\allowdisplaybreaks
\setlength{\jot}{6pt} 
\begin{align} \label{eq:o1-10}
	O_1^u &= \left(\bar{s}_{\alpha}u_{\beta} \right)_{V-A}\left(\bar{u}_{\beta}b_{\alpha}\right)_{V-A}, \nonumber \\ 
	O_2^u &= \left(\bar{s}_{\alpha}u_{\alpha}\right)_{V-A}\left(\bar{u}_{\beta}b_{\beta} \right)_{V-A}, \nonumber \\
	O_3 &= (\bar{s}_{\alpha}b_{\alpha})_{V-A}\sum_{q^\prime}\left(\bar{q^\prime}_{\beta}q^\prime_{\beta} \right)_{V-A}, \nonumber \\
	O_4 &= (\bar{s}_{\alpha}b_{\beta})_{V-A}\sum_{q^\prime} \left(\bar{q^\prime}_{\beta}q^\prime_{\alpha}\right)_{V-A}, \nonumber \\
	O_5 &= (\bar{s}_{\alpha}b_{\alpha})_{V-A}\sum_{q^\prime}\left(\bar{q^\prime}_{\beta}q^\prime_{\beta} \right)_{V+A}, \nonumber \\
	O_6 &= (\bar{s}_{\alpha}b_{\beta})_{V-A}\sum_{q^\prime} \left(\bar{q^\prime}_{\beta}q^\prime_{\alpha}\right)_{V+A}, \nonumber \\
	O_7   &= \dfrac{3}{2}(\bar{s}_{\alpha}b_{\alpha})_{V-A}\sum_{q^\prime}e_{q^\prime}\left(\bar{q^\prime}_{\beta}q^\prime_{\beta} \right)_{V+A}, \nonumber \\
	O_8   &= \dfrac{3}{2}(\bar{s}_{\alpha}b_{\beta})_{V-A}\sum_{q^\prime}e_{q^\prime} \left(\bar{q^\prime}_{\beta}q^\prime_{\alpha}\right)_{V+A}, \nonumber \\
	O_9   &= \dfrac{3}{2}(\bar{s}_{\alpha}b_{\alpha})_{V-A}\sum_{q^\prime}e_{q^\prime}\left(\bar{q^\prime}_{\beta}q^\prime_{\beta} \right)_{V-A}, \nonumber \\
	O_{10}&= \dfrac{3}{2}(\bar{s}_{\alpha}b_{\beta})_{V-A}\sum_{q^\prime}e_{q^\prime} \left(\bar{q^\prime}_{\beta}q^\prime_{\alpha}\right)_{V-A}, \nonumber \\
	O_{8g}&=\frac{g_s}{8\pi^2}m_b\bar{s}_\alpha\sigma^{\mu\nu}(1+\gamma_5)T^a_{\alpha\beta}G^a_{\mu\nu}b_\beta, 
\end{align}
}
where the sum of $q^\prime$ is taken over the set $\{u, d, s, c, b\}$. 
The Greek letters $\alpha$ and $\beta$ denote the color indices, and the subscript $V\pm A$ refers to the Lorentz structure $\gamma_\mu(1\pm \gamma_5)$.

\subsection{Meson distribution amplitudes \label{subsec:lcda}}
In the PQCD approach, the light-cone distribution amplitudes (LCDA) of mesons are important inputs. 
The $B$ meson is treated as a heavy-light system. 
Its distribution amplitude under the heavy quark effective theory is defined by the matrix operator $\langle 0|\bar{q}_{\gamma}(z)[z,0]b_{\delta}(0)|\bar{B}\rangle$ as
\begin{equation}
	\langle 0|\bar{q}_{\gamma}(z)[z,0]b_{\delta}(0)|\bar{B}\rangle = \int d^3k \Phi_{\delta\gamma}^B(k)\exp(-ik\cdot z),
\end{equation}
with $[z,0]$ being the Wilson line and $\gamma$ and $\delta$ the spinor indices. 

In this work, we adopt the $B$-meson distribution amplitude based the QCD-inspired relativistic model \cite{Yang:2011ie, Liu:2013maa, Liu:2015lka}. 
The spinor wave function $\Phi_{\delta\gamma}^B(k)$ is given by \cite{Sun:2016avp, Sun:2019xyw}
\begin{align}
	\Phi_{\delta\gamma}^B(k)&=\frac{-if_Bm_B}{4}K(\vec{k}) \nonumber \\ 
		&\times\biggl\{(E_Q+m_Q)\frac{1+\not{v}}{2}\biggl[\left(\frac{k_+}{\sqrt{2}}+\frac{m_q}{2}\right)\not{n}_+ \nonumber \\
		&\quad+\left(\frac{k_-}{\sqrt{2}}+\frac{m_q}{2}\right)\not{n}_{-}-k_\perp^\mu\gamma_\mu\biggr]\gamma_5 \nonumber \\
		&\quad-(E_q+m_q)\frac{1-\not{v}}{2}\biggl[\left(\frac{k_+}{\sqrt{2}}-\frac{m_q}{2}\right)\not{n}_+ \nonumber \\
		&\quad+\left(\frac{k_-}{\sqrt{2}}-\frac{m_q}{2}\right)\not{n}_--k_\perp^\mu\gamma_\mu\biggr]\gamma_5\biggr\}_{\alpha\beta},
\end{align}
where $f_B$ and $m_B$ are the decay constant and the mass of $B$ meson. 
The subscripts $Q$ and $q$ of the $E$ and $m$ denote the heavy and light quarks within the $B$ meson, respectively. 
The symbol $v$ is the four-speed of $B$ meson, and the lightlike vectors $n_{\pm}^\mu$ are defined by $n_{\pm}^\mu = (1, 0, 0, \mp 1)$. 
The momentum $k$ of the light quark is decomposed in light-cone coordinates as follows: 
\begin{equation}
	k^\pm=\frac{E_q\pm k^3}{\sqrt{2}},\quad k_\perp^\mu=(0,k^1,k^2,0).
\end{equation}
The function $K(\vec{k})$ is proportional to the $B$-meson wave function 
\begin{align}
	K(\vec{k}) &= \frac{2N_B\Psi_0(\vec{k})}{\sqrt{E_qE_Q(E_q+m_q)(E_Q+m_Q)}} \nonumber \\ 
		&= \frac{2N_Ba_1 \exp\left(a_2 |\vec{k}|^2 + a_3 |\vec{k}| +a_4\right)}{\sqrt{E_qE_Q(E_q+m_q)(E_Q+m_Q)}},
\end{align}
with the normalization factor $N_B=\sqrt{{3}/\left[(2\pi)^3m_B\right]}/f_B$ and the parameters $a_{1-4}$ given by \cite{Sun:2016avp}
\begin{align} \label{eq:a1234}
	a_1 &= 4.55_{-0.30}^{+0.40}~\text{GeV}^{-3/2}, & a_2 &= -0.39_{-0.20}^{+0.15}~\text{GeV}^{-2}, \nonumber\\
    a_3 &= -1.55\pm 0.20~\text{GeV}^{-1}, & a_4 &= -1.10_{-0.05}^{+0.10}.
\end{align}

To investigate the $B\to K^*\pi$ and $B\to K\rho$ decays, the LCDAs for vector meson ($K^*$ and $\rho$) and pseudoscalar meson ($\pi$ and $K$) are also needed. 
The $B$ meson is much heavier than the final state mesons. 
Therefore, it is reasonable to define the wave functions of large-recoiling final state mesons on the light cone. 
For the pseudoscalar meson, we take $K^-$ meson as an example, from which the wave functions of other mesons can be obtained by simple substitutions. 
Up to twist-3 accuracy, the vacuum to kaon matrix element can be expressed in terms of the twist expansion as follows \cite{Braun:1989iv, Ball:1998je, Ball:2006wn}: 
\begin{align}
	\langle & K^-(p)|\bar{s}_\xi(y)u_\eta(0)|0\rangle \nonumber \\
	&=\int dx d^2k_{q\perp} \exp\left[i(xp\cdot y - k_{q\perp}\cdot y_\perp)\right] \Phi_{\eta\xi}^K(x,k_{q\perp}) \nonumber \\ 
	&=\dfrac{i}{\sqrt{2N_c}}\int dx d^2k_{q\perp} \exp\left[i(xp\cdot y - k_{q\perp}\cdot y_\perp)\right] \nonumber \\
	&\quad \times \left[\not{p}\gamma_5\phi_K^A(x,k_{q\perp})-\mu_K\gamma_5\phi_K^P(x,k_{q\perp}) \right. \nonumber \\
	&\quad \left. +\mu_K\gamma_5\sigma^{\mu\nu}p_\mu z_\nu \phi_K^T(x,k_{q\perp})\right]_{\eta\xi} \nonumber \\
	&\to \dfrac{i}{\sqrt{2N_c}}\int dx d^2k_{q\perp} \exp\left[i(xp\cdot y - k_{q\perp}\cdot y_\perp)\right] \nonumber \\
	&\quad \times \left[\not{p}\gamma_5\phi_K^A(x,k_{q\perp})-\mu_K\gamma_5\phi_K^P(x,k_{q\perp}) \right. \nonumber \\
	&\quad \left. +i\mu_K\gamma_5\sigma^{\mu\nu}\left(p_\mu \bar{p}_\nu/(p\cdot\bar{p})\right) \phi_K^T(x,k_{q\perp})\right]_{\eta\xi}, \label{k-wave}
\end{align}
where $N_c=3$ is the number of colors, the $\mu_K=m_K^2/(m_s+m_u)$ is the chiral mass, and $\bar{p}$ is the momentum with the opposite direction of $p$. 
Note that $k_{q\perp}$ is used to represent the transverse momentum of the quark in the $K^-$ meson. 
In the last step, we disregard the term involving the derivative of transverse momentum. 
The detailed expressions of the LCDAs $\phi_K^{A,P,T}$ are shown in Appendix.~\ref{app:lightda}. 

For the vector meson, it is essential to notice that polarization should be taken into consideration when defining the LCDAs. 
Here, the final state vector meson in longitudinal polarization is the only situation we will encounter. 
Similarly, up to twist-3 accuracy, the LCDAs of the longitudinal polarized $K^*$ meson are defined by the matrix element as \cite{Ball:1996tb, Ball:1998sk, Ball:2007rt,Kurimoto:2001zj}
\begin{align}
	\langle & K^{*-}(p,\epsilon_L)|\bar{s}_\xi(y)u_\eta(0)|0\rangle \nonumber \\
	&=\int dx d^2k_{q\perp} \exp\left[i(xp\cdot y - k_{q\perp}\cdot y_\perp)\right] \Phi_{\eta\xi}^{K^*}(x,k_{q\perp}) \nonumber \\ 
	&=\dfrac{1}{\sqrt{2N_c}}\int dx d^2k_{q\perp} \exp\left[i(xp\cdot y - k_{q\perp}\cdot y_\perp)\right] \nonumber \\
	&\quad \times\left[m_{K^*}\not{\epsilon_L}\phi_{K^*}(x,k_{q\perp}) + \not{\epsilon_L}\not{p}\phi_{K^*}^t(x,k_{q\perp})\right. \nonumber \\ 
	&\qquad + \left. m_{K^*}\phi_{K^*}^s(x,k_{q\perp})\right]_{\eta\xi}, 
\end{align}
where $m_{K^*}$ is the mass of $K^*$ meson, $\epsilon_L$ denotes the longitudinal polarization vector. 
The twist-2 and twist-3 LCDAs, $\phi_{K^*}$ and $\phi_{K^*}^{s,t}$, are also arranged in Appendix.~\ref{app:lightda}. 

\subsection{Leading order decay amplitudes \label{subsec:loamp}}
\begin{figure*}
	\includegraphics[width=0.8\textwidth]{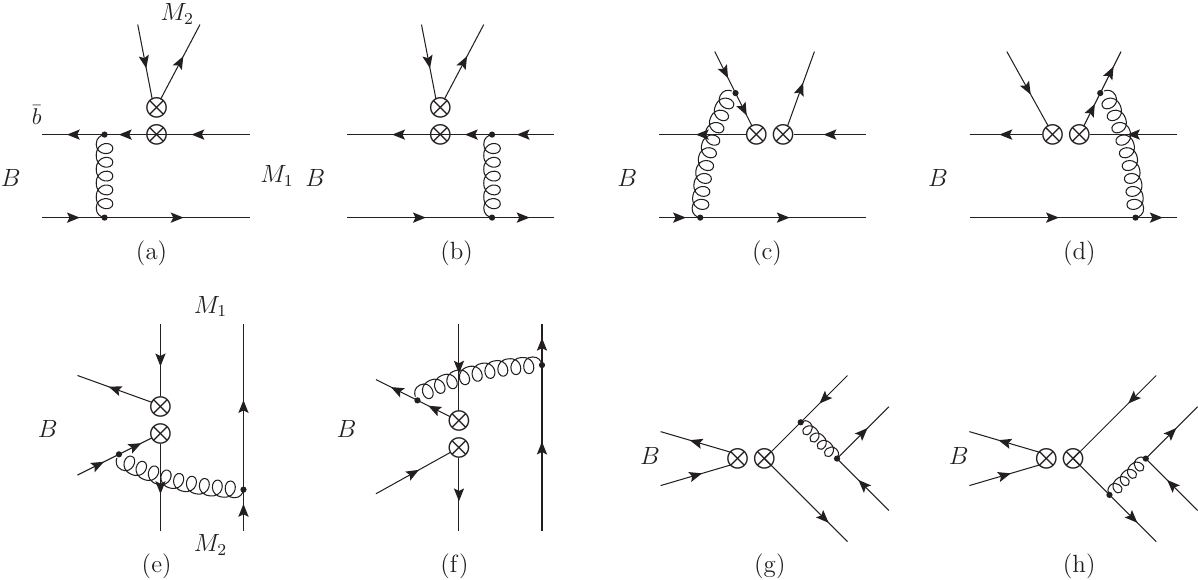}
	\caption{\label{fig:eightdiagram} Decay diagrams contributing to the $B\to K^*\pi$ and $B\to K\rho$ decays. 
		Diagrams (a, b) are factorizable emission diagrams, (c, d) are nonfactorizable emission diagrams, 
		(e, f) are nonfactorizable annihilation diagrams, and (g, h) are factorizable annihilation diagrams.}
\end{figure*}

At leading order, the diagrams which contribute to the $B\to K^*\pi$ and $B\to K\rho$ decays are shown in Fig.~\ref{fig:eightdiagram}. 
These are eight diagrams classified as four types: factorizable emission diagrams (Fig.~\ref{fig:eightdiagram}(a) and (b)), 
nonfactorizable emission diagrams (Fig.~\ref{fig:eightdiagram}(c) and (d)), nonfactorizable annihilation diagrams (Fig.~\ref{fig:eightdiagram}(e) and (f)), 
and factorizable annihilation diagrams (Fig.~\ref{fig:eightdiagram}(g) and (h)). 
The pair of cross circle symbols $\otimes$ in the diagrams represent the insertion of four-quark operators. 
As listed in Eq.~\eqref{eq:o1-10}, there are three kinds of possible Lorentz structure for the four-quark operators: 
$(V-A)(V-A)$, $(V-A)(V+A)$, and $(S+P)(S-P)$ which is derived from the Fierz transformation. 

Based on the type of diagrams and the inserted four-quark operators, we denote the results of these diagrams as follows: 
$F_e,\ F_e^R,\ F_e^P$, $M_e,\ M_e^R,\ M_e^P$, $F_a,\ F_a^R,\ F_a^P$, $M_a,\ M_a^R,\ M_a^P$, 
where $F$ and $M$ are used to distinguish factorizable and nonfactorizable diagram, the subscript is employed for emission or annihilation diagrams, 
and the superscript indicates the type of four-quark operators.
Moreover, the final state mesons have two possible configurations. 
For instance, in the $B\to K^*\pi$ decay, the meson 1($M_1$) in the diagrams may correspond to either the $K^*$ or the $\pi$ meson. 
We will use the subscripts to identify the meson configuration in the diagrams, such as $F_{e, M_1M_2}^R$.

With the thorough description provided above, the results of these eight diagrams in the PQCD framework are ready to be presented as: 
\begin{widetext}
{
\allowdisplaybreaks 
\setlength{\jot}{10pt} 
\begin{align} \label{eq:fe}
	F_{e,\pi K^*}=&-4\sqrt{2N_c}i\pi^2\frac{C_F}{N_c}f_Bf_{K^*}^{\parallel}m_B^2 \int k_{0\perp}dk_{0\perp}\int_{x_0^d}^{x_0^u}dx_0\int_0^1dx_1 \int_0^\infty b_0db_0b_1db_1
			\left(\frac{m_B}{2}+\frac{|\mathbf{k}_{0\perp}|^2}{2x_0^2m_B}\right) K(\vec{k}_0)(E_Q+m_Q) \nonumber \\
		&\times J_0(k_{0\perp}b_0)\{\alpha_s(t_e^1)[((x_1-2)E_q-x_1k_0^3)\phi_{\pi}^A(x_1,b_1)+r_{\pi}((1-2x_1)E_q+k_0^3)(\phi_{\pi}^P(x_1,b_1)-\phi_{\pi}^T(x_1,b_1))] \nonumber \\
		&\times h_e(x_0,1-x_1,b_0,b_1)S_t(x_1)\exp[-S_B(t_e^1)-S_{\pi}(t_e^1)]+\alpha_s(t_e^2)2r_{\pi}(-E_q+k_0^3)\phi_{\pi}^P(x_1,b_1)h_e(1-x_1,x_0,b_1,b_0) \nonumber \\ 
		&\times  S_t(x_0)\exp[-S_B(t_e^2)-S_{\pi}(t_e^2)]\}, 
\end{align}
\begin{equation} \label{eq:ferandfep}
    F_{e,\pi K^*}^R=F_{e,\pi K^*}, \quad F_{e,\pi K^*}^P=0,
\end{equation}
\begin{align} \label{eq:fep}
	F_{e,K^*\pi}^P=&4\sqrt{2N_c}i\pi^2\frac{C_F}{N_c}f_Bf_{\pi}m_B^2\int k_{0\perp}dk_{0\perp}\int_{x_0^d}^{x_0^u}dx_0\int_0^1dx_1\int_0^\infty b_0db_0b_1db_1
			\left(\frac{m_B}{2}+\frac{|\mathbf{k}_{0\perp}|^2}{2x_0^2m_B}\right)K(\vec{k}_0)(E_Q+m_Q) \nonumber \\
		&\times J_0(k_{0\perp}b_0)2r_{\pi}\{\alpha_s(t_e^1)\left[-(E_q-k_0^3)\phi_{K^*}(x_1,b_1)-r_{K^*}((x_1-3)E_q-(1-x_1)k_0^3)\phi_{K^*}^s(x_1,b_1)\right. \nonumber \\ 
		& \left.-r_{K^*}((x_1-1)E_q -(1+x_1)k_0^3)\phi_{K^*}^t(x_1,b_1)\right]h_e(x_0,1-x_1,b_0,b_1)S_t(x_1)\exp[-S_B(t_e^1)-S_{K^*}(t_e^1)] \nonumber \\
		&+\alpha_s(t_e^2)2r_{K^*}(E_q+k_0^3)\phi_{K^*}^s(x_1,b_1)h_e(1-x_1,x_0,b_1,b_0)S_t(x_0)\exp[-S_B(t_e^2)-S_{K^*}(t_e^2)]\}, 
\end{align}
\begin{align} \label{eq:me}
	\mathcal{M}_{e,\pi K^*}=&-16i\pi^2C_Ff_Bm_B^2 \int k_{0\perp}dk_{0\perp}\int_{x_0^d}^{x_0^u}dx_0\int_0^1dx_1dx_2 \int_0^\infty b_0db_0b_2db_2
			\left(\frac{m_B}{2}+\frac{|\mathbf{k}_{0\perp}|^2}{2x_0^2m_B}\right) K(\vec{k}_0)(E_Q+m_Q) \nonumber \\ 
		&\times J_0(k_{0\perp}b_0)\phi_{K^*}(x_2,b_2) \{\alpha_s(t_{ne}^1) [-x_2(E_q-k_0^3)\phi_{\pi}^A(x_1,b_0)+r_{\pi}(1-x_1)(E_q+k_0^3)(\phi_{\pi}^P(x_1,b_0) \nonumber \\
		&+\phi_{\pi}^T(x_1,b_0))]h_{ne}(x_0,x_2,1-x_1,b_0,b_2)\exp[-S_B(t_{ne}^1)-S_{\pi}(t_{ne}^1)|_{b_1\rightarrow b_0}-S_{K^*}(t_{ne}^1)]+\alpha_s(t_{ne}^2) \nonumber \\ 
		&\times [((2-x_1-x_2)E_q-(x_1-x_2)k_0^3)\phi_{\pi}^A(x_1,b_0)-r_{\pi}(1-x_1)(E_q-k_0^3)(\phi_{\pi}^P(x_1,b_0)-\phi_{\pi}^T(x_1,b_0))] \nonumber \\ 
		&\times h_{ne}(x_0,1-x_2,1-x_1,b_0,b_2)\exp[-S_B(t_{ne}^2)-S_{\pi}(t_{ne}^2)|_{b_1\rightarrow b_0}-S_{K^*}(t_{ne}^2)]\},
\end{align}
\begin{align} \label{eq:mer}
	\mathcal{M}_{e,\pi K^*}^R=&16i\pi^2C_Ff_Bm_B^2\int k_{0\perp}dk_{0\perp}\int_{x_0^d}^{x_0^u}dx_0\int_0^1dx_1dx_2\int_0^\infty b_0db_0b_2db_2
			\left(\frac{m_B}{2}+\frac{|\mathbf{k}_{0\perp}|^2}{2x_0^2m_B}\right)K(\vec{k}_0)(E_Q+m_Q) \nonumber \\
		&\times J_0(k_{0\perp}b_0)r_{K^*}\{\alpha_s(t_{ne}^1)[x_2(E_q-k_0^3)\phi_{\pi}^A(x_1,b_0)(-\phi_{K^*}^s(x_2,b_2)+\phi_{K^*}^t(x_2,b_2)) \nonumber \\ 
		&+r_{\pi}((1-x_1+x_2)E_q+(1-x_1-x_2)k_0^3)(-\phi_{\pi}^P(x_1,b_0)\phi_{K^*}^s(x_2,b_2)+\phi_{\pi}^T(x_1,b_0)\phi_{K^*}^t(x_2,b_2)) \nonumber \\ 
		&+r_{\pi}((1-x_1-x_2)E_q+(1-x_1+x_2)k_0^3)(\phi_{\pi}^T(x_1,b_0)\phi_{K^*}^s(x_2,b_2)-\phi_{\pi}^P(x_1,b_0)\phi_{K^*}^t(x_2,b_2))] \nonumber\\ 
		&\times h_{ne}(x_0,x_2,1-x_1,b_0,b_2)\exp[-S_B(t_{ne}^1)-S_{\pi}(t_{ne}^1)|_{b_1\rightarrow b_0}-S_{K^*}(t_{ne}^1)] \nonumber \\ 
		&+\alpha_s(t_{ne}^2)[(1-x_2)(E_q-k_0^3)\phi_{\pi}^A(x_1,b_0)(\phi_{K^*}^s(x_2,b_2)+\phi_{K^*}^t(x_2,b_2)) \nonumber \\ 
		&+r_{\pi}((2-x_1-x_2)E_q-(x_1-x_2)k_0^3)(\phi_{\pi}^P(x_1,b_0)\phi_{K^*}^s(x_2,b_2)+\phi_{\pi}^T(x_1,b_0)\phi_{K^*}^t(x_2,b_2)) \nonumber \\ 
		&+r_{\pi}((x_1-x_2)E_q-(2-x_1-x_2)k_0^3)(\phi_{\pi}^T(x_1,b_0)\phi_{K^*}^s(x_2,b_2)+\phi_{\pi}^P(x_1,b_0)\phi_{K^*}^t(x_2,b_2))] \nonumber \\ 
		&\times h_{ne}(x_0,1-x_2,1-x_1,b_0,b_2)\exp[-S_B(t_{ne}^2)-S_{\pi}(t_{ne}^2)|_{b_1\rightarrow b_0}-S_{K^*}(t_{ne}^2)]\},
\end{align}
\begin{align} \label{eq:mep}
	\mathcal{M}_{e,\pi K^*}^P=&16i\pi^2C_Ff_Bm_B^2\int k_{0\perp}dk_{0\perp}\int_{x_0^d}^{x_0^u}dx_0\int_0^1dx_1dx_2\int_0^\infty b_0db_0b_2db_2
			\left(\frac{m_B}{2}+\frac{|\mathbf{k}_{0\perp}|^2}{2x_0^2m_B}\right)K(\vec{k}_0)(E_Q+m_Q) \nonumber \\ 
		&\times J_0(k_{0\perp}b_0)\phi_{K^*}(x_2,b_2)\{\alpha_s(t_{ne}^1)[((1-x_1+x_2)E_q+(1-x_1-x_2)k_0^3)\phi_{\pi}^A(x_1,b_0) \nonumber \\ 
		&-r_{\pi}(1-x_1)(E_q-k_0^3)(\phi_{\pi}^P(x_1,b_0)-\phi_{\pi}^T(x_1,b_0))]h_{ne}(x_0,x_2,1-x_1,b_0,b_2) \nonumber \\ 
		&\times\exp[-S_B(t_{ne}^1)-S_{\pi}(t_{ne}^1)|_{b_1\rightarrow b_0}-S_{K^*}(t_{ne}^1)]+\alpha_s(t_{ne}^2)[-(1-x_2)(E_q-k_0^3)\phi_{\pi}^A(x_1,b_0) \nonumber \\ 
		&+r_{\pi}(1-x_1)(E_q+k_0^3)(\phi_{\pi}^P(x_1,b_0)+\phi_{\pi}^T(x_1,b_0))]h_{ne}(x_0,1-x_2,1-x_1,b_0,b_2) \nonumber \\ 
		&\times\exp[-S_B(t_{ne}^2)-S_{\pi}(t_{ne}^2)|_{b_1\rightarrow b_0}-S_{K^*}(t_{ne}^2)]\},
\end{align}
\begin{align} \label{eq:fa}
	F_{a,\pi K^*}=&-8i\pi C_Ff_Bm_B^2 \int_0^1dx_1dx_2\int_0^\infty b_1db_1b_2db_22\{\alpha_s(t_a^1)\left[x_2\phi_{\pi}^A(x_1,b_1)\phi_{K^*}(x_2,b_2)+2r_{\pi}r_{K^*}(1+x_2)\right. \nonumber\\ 
		&\left.\times\phi_{\pi}^P(x_1,b_1)\phi_{K^*}^s(x_2,b_2)-2r_{\pi}r_{K^*}(1-x_2)\phi_{\pi}^P(x_1,b_1)\phi_{K^*}^t(x_2,b_2)\right]h_a(1-x_1,x_2,b_1,b_2)S_t(x_2) \nonumber\\ 
		&\times\exp[-S_{\pi}(t_a^1)-S_{K^*}(t_a^1)]+\alpha_s(t_a^2)[-(1-x_1)\phi_{\pi}^A(x_1,b_1)\phi_{K^*}(x_2,b_2)-2r_{\pi}r_{K^*}(2-x_1) \nonumber\\ 
		&\times\phi_{\pi}^P(x_1,b_1)\phi_{K^*}^s(x_2,b_2)-2r_{\pi}r_{K^*}x_1\phi_{\pi}^T(x_1,b_1)\phi_{K^*}^s(x_2,b_2)]h_a(x_2,1-x_1,b_2,b_1)S_t(x_1) \nonumber\\
		&\times\exp[-S_{\pi}(t_a^2)-S_{K^*}(t_a^2)]\},
\end{align}
\begin{equation} \label{eq:far}
	F_{a,\pi K^*}^R = -F_{a,\pi K^*},
\end{equation}
\begin{align} \label{eq:fap}
	F_{a,\pi K^*}^P=&8i\pi C_Ff_Bm_B^2\chi_B\int_0^1dx_1dx_2\int_0^\infty b_1db_1b_2db_2 4\{\alpha_s(t_a^1)[-r_{K^*}x_2\phi_{\pi}^A(x_1,b_1)(\phi_{K^*}^s(x_2,b_2)-\phi_{K^*}^t(x_2,b_2)) \nonumber\\
		&-2r_{\pi}\phi_{\pi}^P(x_1,b_1)\phi_{K^*}(x_2,b_2)]h_a(1-x_1,x_2,b_1,b_2)S_t(x_2)\exp[-S_{\pi}(t_a^1)-S_{K^*}(t_a^1)]+\alpha_s(t_a^2) \nonumber \\ 
		&\times[-2r_{K^*}\phi_{\pi}^A(x_1,b_1)\phi_{K^*}^s(x_2,b_2)-r_{\pi}(1-x_1)(\phi_{\pi}^P(x_1,b_1)+\phi_{\pi}^T(x_1,b_1))\phi_{K^*}(x_2,b_2)] \nonumber \\
		&\times h_a(x_2,1-x_1,b_2,b_1)S_t(x_1)\exp[-S_{\pi}(t_a^2)-S_{K^*}(t_a^2)]\}, 
\end{align}
\begin{align} \label{eq:ma}
	\mathcal{M}_{a,\pi K^*}=&-16i\pi^2C_Ff_Bm_B^2\int k_{0\perp}dk_{0\perp}\int_{x_0^d}^{x_0^u}dx_0\int_0^1dx_1dx_2\int_0^\infty b_0db_0b_1db_1
			\left(\frac{m_B}{2}+\frac{|\mathbf{k}_{0\perp}|^2}{2x_0^2m_B}\right)K(\vec{k}_0)(E_Q+m_Q) \nonumber \\ 
		&\times J_0(k_{0\perp}b_0)\{\alpha_s(t_{na}^1)[(1-x_1)(E_q+k_0^3)\phi_{\pi}^A(x_1,b_1)\phi_{K^*}(x_2,b_1)
			+r_{\pi}r_{K^*}((3-x_1+x_2)E_q \nonumber \\ &+(1-x_1-x_2)k_0^3)\phi_{\pi}^P(x_1,b_1)\phi_{K^*}^s(x_2,b_1) 
			+r_{\pi}r_{K^*}((1-x_1-x_2)E_q-(1+x_1-x_2)k_0^3) \nonumber \\ &\times\phi_{\pi}^P(x_1,b_1)\phi_{K^*}^t(x_2,b_1)
			-r_{\pi}r_{K^*}((1-x_1-x_2)E_q+(3-x_1+x_2)k_0^3)\phi_{\pi}^T(x_1,b_1)\phi_{K^*}^s(x_2,b_1) \nonumber \\ 
		&+r_{\pi}r_{K^*}((1+x_1-x_2)E_q-(1-x_1-x_2)k_0^3)\phi_{\pi}^T(x_1,b_1)\phi_{K^*}^t(x_2,b_1)]h_{na}^1(1-x_1,x_2,b_0,b_1) \nonumber \\ 
		&\times \exp[-S_B(t_{na}^1)-S_{\pi}(t_{na}^1)-S_{K^*}(t_{na}^1)|_{b_2\rightarrow b_1}]+\alpha_s(t_{na}^2)[-x_2(E_q-k_0^3)\phi_{\pi}^A(x_1,b_1)\phi_{K^*}(x_2,b_1) \nonumber \\ 
		&-r_{\pi}r_{K^*}((1-x_1+x_2)E_q+(1-x_1-x_2)k_0^3)(\phi_{\pi}^P(x_1,b_1)\phi_{K^*}^s(x_2,b_1)-\phi_{\pi}^T(x_1,b_1)\phi_{K^*}^t(x_2,b_1)) \nonumber \\ 
		&+r_{\pi}r_{K^*}((1-x_1-x_2)E_q+(1-x_1+x_2)k_0^3)(\phi_{\pi}^P(x_1,b_1)\phi_{K^*}^t(x_2,b_1)-\phi_{\pi}^T(x_1,b_1)\phi_{K^*}^s(x_2,b_1))] \nonumber \\ 
		&\times h_{na}^2(1-x_1,x_2,b_0,b_1)\exp[-S_B(t_{na}^2)-S_{\pi}(t_{na}^2)-S_{K^*}(t_{na}^2)|_{b_2\rightarrow b_1}]\}, 
\end{align}
\begin{align} \label{eq:mar}
	\mathcal{M}_{a,\pi K^*}^R=&16i\pi^2C_Ff_Bm_B^2\int k_{0\perp}dk_{0\perp}\int_{x_0^d}^{x_0^u}dx_0\int_0^1dx_1dx_2\int_0^\infty b_0db_0b_1db_1
			\left(\frac{m_B}{2}+\frac{|\mathbf{k}_{0\perp}|^2}{2x_0^2m_B}\right)K(\vec{k}_0)(E_Q+m_Q) \nonumber \\ 
		&\times J_0(k_{0\perp}b_0)\{\alpha_s(t_{na}^1)[r_{K^*}((2-x_2)E_q+x_2k_0^3)\phi_{\pi}^A(x_1,b_1)(\phi_{K^*}^s(x_2,b_1)+\phi_{K^*}^t(x_2,b_1)) \nonumber \\ 
		&-r_{\pi}((1+x_1)E_q-(1-x_1)k_0^3)(\phi_{\pi}^P(x_1,b_1)-\phi_{\pi}^T(x_1,b_1))\phi_{K^*}(x_2,b_1)]h_{na}^1(1-x_1,x_2,b_0,b_1) \nonumber \\ 
		&\times \exp[-S_B(t_{na}^1)-S_{\pi}(t_{na}^1)-S_{K^*}(t_{na}^1)|_{b_2\rightarrow b_1}]+\alpha_s(t_{na}^2)
			[r_{K^*}x_2(E_q+k_0^3)\phi_{\pi}^A(x_1,b_1)(\phi_{K^*}^s(x_2,b_1) \nonumber \\ & +\phi_{K^*}^t(x_2,b_1))
			-r_{\pi}(1-x_1)(E_q-k_0^3)(\phi_{\pi}^P(x_1,b_1)-\phi_{\pi}^T(x_1,b_1))\phi_{K^*}(x_2,b_1)]h_{na}^2(1-x_2,x_3,b_1,b_2) \nonumber \\ 
		&\times\exp[-S_B(t_{na}^2)-S_{\pi}(t_{na}^2)-S_{K^*}(t_{na}^2)|_{b_2\rightarrow b_1}]\}, 
\end{align}
\begin{align} \label{eq:map}
	\mathcal{M}_{a,\pi K^*}^P=&16i\pi^2C_Ff_Bm_B^2\int k_{0\perp}dk_{0\perp}\int_{x_0^d}^{x_0^u}dx_0\int_0^1dx_1dx_2\int_0^\infty b_0db_0b_1db_1
			\left(\frac{m_B}{2}+\frac{|\mathbf{k}_{0\perp}|^2}{2x_0^2m_B}\right)K(\vec{k}_0)(E_Q+m_Q) \nonumber \\ 
		&\times J_0(k_{0\perp}b_0)\{\alpha_s(t_{na}^1)[x_2(E_q-k_0^3)\phi_{\pi}^A(x_1,b_1)\phi_{K^*}(x_2,b_1)
			+r_{\pi}r_{K^*}((3-x_1+x_2)E_q \nonumber \\ & +(1-x_1-x_2)k_0^3)\phi_{\pi}^P(x_1,b_1)\phi_{K^*}^s(x_2,b_1) 
			-r_{\pi}r_{K^*}((1-x_1-x_2)E_q+(3-x_1+x_2)k_0^3) \nonumber \\ &\times \phi_{\pi}^P(x_1,b_1)\phi_{K^*}^t(x_2,b_1)
			+r_{\pi}r_{K^*}((1-x_1-x_2)E_q-(1+x_1-x_2)k_0^3)\phi_{\pi}^T(x_1,b_1)\phi_{K^*}^s(x_2,b_1) \nonumber \\ 
		&+r_{\pi}r_{K^*}((1+x_1-x_2)E_q-(1-x_1-x_2)k_0^3)\phi_{\pi}^T(x_1,b_1)\phi_{K^*}^t(x_2,b_1)]h_{na}^1(1-x_1,x_2,b_0,b_1) \nonumber \\ 
		&\times \exp[-S_B(t_{na}^1)-S_{\pi}(t_{na}^1)-S_{K^*}(t_{na}^1)|_{b_2\rightarrow b_1}]
			+\alpha_s(t_{na}^2)[-(1-x_1)(E_q+k_0^3)\phi_{\pi}^A(x_1,b_1)\phi_{K^*}(x_2,b_1) \nonumber \\ 
		&-r_{\pi}r_{K^*}((1-x_1+x_2)E_q+(1-x_1-x_2)k_0^3)(\phi_{\pi}^P(x_1,b_1)\phi_{K^*}^s(x_2,b_1)-\phi_{\pi}^T(x_1,b_1)\phi_{K^*}^t(x_2,b_1)) \nonumber \\ 
		&-r_{\pi}r_{K^*}((1-x_1-x_2)E_q+(1-x_1+x_2)k_0^3)(\phi_{\pi}^P(x_1,b_1)\phi_{K^*}^t(x_2,b_1)-\phi_{\pi}^T(x_1,b_1)\phi_{K^*}^s(x_2,b_1))] \nonumber \\ 
		&\times h_{na}^2(1-x_1,x_2,b_0,b_1)\exp[-S_B(t_{na}^2)-S_{\pi}(t_{na}^2)-S_{K^*}(t_{na}^2)|_{b_2\rightarrow b_1}]\}.
\end{align}
}
\end{widetext}
There are some symbols needed to be explained in Eqs.~\eqref{eq:fe}-\eqref{eq:map}. 
$C_F = (N_c^2-1)/2N_c$ is the color factor. 
$f_{K^*}^\parallel$ and $f_{\pi}$ are decay constants of the $K^*$ and $\pi$ mesons. 
$\alpha_s(t)$ is the strong running coupling at the scale of $t$. 
$r_{K^*}$ and $r_{\pi}$ are dimensionless ratios of chiral mass and $B$ meson mass. 
They are defined by 
\begin{align}
	r_P &= m_P^2/[(m_{q1}+m_{q2})m_B],\quad \text{for pseudoscalar meson} \nonumber \\ 
	r_V &= m_V/m_B,\quad \text{for vector meson}.
\end{align}
The function $J_0$ is the Bessel function of the first kind, and the hard function $h$'s are also consisted of the Bessel functions. 
The factor $\chi_B$ in Eq.~\eqref{eq:fap} used for normalization is defined by 
\begin{align} \label{eq:chib}
	\chi_B &= \pi\int dk_{0\perp}k_{0\perp}\int_{x_0^d}^{x_0^u}dx_0 \left(\frac{m_B}{2}+\frac{|\mathbf{k}_{0\perp}|^2}{2x_0^2m_B}\right) \nonumber \\
	  &\quad \times K(\vec{k}_0)\left[(E_q+m_q)(E_Q+m_Q)+|\vec{k}_0|^2\right] \nonumber \\
	  &\approx 1.388.
\end{align}
The Sudakov factor $S_M(t)$ and the threshold factor $S_t(x)$ are derived through the technique of resummation. 
The detailed expressions of hard function $h$'s, $t$ scales, the Sudakov factor, and the threshold factor are all collected in Appendix.~\ref{app:sudakovht}. 

As for the situation that the meson 1 ($M_1$) being $K^*$ and $M_2$ being pion, 
the results of these diagrams have the same form as previously and can be obtained by a straightforward substitution: 
$\{f_\pi\leftrightarrow f_{K^*}$, $r_\pi \leftrightarrow -r_{K^*}$, $\phi_\pi^{A,P,T}\leftrightarrow \phi_{K^*},\phi_{K^*}^{s,t}\}$. 
The only exception is the result of the factorizable emission diagrams with $(S+P)(S-P)$ operator inserted. 
Due to the different parities of the $\pi$ and $K^*$ mesons, the matrix elements of the meson to the vacuum with the insertion of $(S+P)(S-P)$ operator yield contrasting results: 
for the $K^*$ meson, the result is zero, while for the $\pi$ meson, it is nonzero. 
That is the reason that we list the result $F_{e,K^*\pi}^P$ in Eq.~\eqref{eq:fep} in addition to the $F_{e,\pi K^*}^P$. 
Furthermore, we can replace the meson parameters to derive the results for the $B\to K\rho$ decays. 

The next step is organizing the results of the eight diagrams with the corresponding Wilson coefficients to calculate the complete amplitude for each decay channel. 
Here, our discussion involves eight decay channels. 
For simplicity, we arrange the Wilson coefficients as
\begin{align} \label{eq:Wilsona}
	a_1(\mu)=&C_2(\mu)+\frac{C_1(\mu)}{N_c},\quad a_2(\mu)=C_1(\mu)+\frac{C_2(\mu)}{N_c}, \nonumber \\
    a_i(\mu)=&C_i(\mu)+\frac{C_{i+1}(\mu)}{N_c},\ a_j(\mu)=C_j(\mu)+\frac{C_{j-1}(\mu)}{N_c}, 
\end{align}
with $i=3,5,7,9$ and $j=4,6,8,10$. 
The decay amplitudes of the $B\to K^*\pi$ and $B\to K\rho$ channels can be expressed in terms of $F$'s and $\mathcal{M}$'s as follows: 
\begin{widetext}
\allowdisplaybreaks 
\setlength{\jot}{10pt} 
\begin{align} 
    \mathcal{M}&(B^-\to \bar{K}^{*0}\pi^-) 
		=V_u\left[a_1F_{a,K^*\pi}+\left(C_1/N_c\right)\mathcal{M}_{a,K^* \pi}\right]-V_t\left\{\left(a_4-a_{10}/2\right)F_{e,\pi K^*}+\left(a_6-a_8/2\right)F_{e,\pi K^*}^P \right. \nonumber \\ 
		&+\left(a_4+a_{10}\right)F_{a,K^*\pi} +\left(a_6+a_8\right)F_{a,K^*\pi}^P+{1}/{N_c}\left[\left(C_3-C_9/2\right)\mathcal{M}_{e,\pi K^*}+\left(C_5-C_7/2\right)\mathcal{M}_{e,\pi K^*}^R \right.\nonumber \\
		&\left.\left. +(C_3+C_9)\mathcal{M}_{a,K^* \pi}+\left(C_5+C_7\right)\mathcal{M}_{a,K^*\pi}^R\right]\right\}, \label{eq:KstarPiamp} \\
	\sqrt{2}\mathcal{M}&(B^-\to K^{*-}\pi^0)
		=V_u\left[a_1F_{e,\pi K^*}+a_2F_{e,K^*\pi}+a_1F_{a,K^*\pi}+1/N_c\left(C_1\mathcal{M}_{e,\pi K^*}+C_2\mathcal{M}_{e,K^*\pi}+C_1\mathcal{M}_{a,K^*\pi}\right)\right] \nonumber \\
		&-V_t\left\{\left(a_4+a_{10}\right)F_{e,\pi K^*} +\left(a_6+a_8\right)F_{e,\pi K^*}^P
			+\left(3a_9/2\right)F_{e,K^*\pi}+\left(3a_7/2\right)F_{e,K^*\pi}^R +\left(a_4+a_{10}\right)F_{a,K^*\pi}\right. \nonumber \\ 
		&+\left(a_6+a_8\right)F_{a,K^*\pi}^P+1/N_c\left[(C_3+C_9)\mathcal{M}_{e,\pi K^*}+(C_5+C_7)\mathcal{M}_{e,\pi K^*}^R
			+\left(3C_{10}/2\right)\mathcal{M}_{e,K^*\pi}+\left(3C_8/2\right)\mathcal{M}_{e,K^*\pi}^P \right. \nonumber \\
		&\left.+(C_3+C_9)\mathcal{M}_{a,K^*\pi}+(C_5+C_7)\mathcal{M}_{a,K^*\pi}^R\right\}, \\
	\mathcal{M}&(\bar{B}^0\to K^{*-}\pi^+) 
		=V_u\left[a_1F_{e,\pi K^*}+\left(C_1/N_c\right)\mathcal{M}_{e,\pi K^*}\right]
			-V_t\left\{\left(a_4+a_{10}\right)F_{e,\pi K^*}+\left(a_6+a_8\right)F_{e,\pi K^*}^P \right. \nonumber\\ 
		&+\left(a_4-a_{10}/2\right)F_{a,K^*\pi}+\left(a_6-a_8/2\right)F_{a,K^*\pi}^P
			+1/N_c\left[(C_3+C_9)\mathcal{M}_{e,\pi K^*}+(C_5+C_7)\mathcal{M}_{e,\pi K^*}^R \right. \nonumber \\ 
		&\left.\left.+(C_3-C_9/2)\mathcal{M}_{a,K^*\pi}+(C_5-C_7/2)\mathcal{M}_{a,K^*\pi}^R\right]\right\}, \\
	-\sqrt{2}\mathcal{M}&(\bar{B}^0\to \bar{K}^{*0}\pi^0) 
		=V_u\left[-a_2F_{e,K^*\pi}-\left(C_2/N_c\right)\mathcal{M}_{e,K^*\pi}\right] 
			-V_t\left\{\left(a_4-a_{10}/2\right)F_{e,\pi K^*}+\left(a_6-a_8/2\right)F_{e,\pi K^*}^P \right. \nonumber \\
		&-\left(3a_9/2\right)F_{e,K^*\pi}-\left(3a_7/2\right)F_{e,K^*\pi}^R +\left(a_4-a_{10}/2\right)F_{a,K^*\pi} +\left(a_6-a_8/2\right)F_{a,K^*\pi}^P \nonumber \\
		&+ 1/N_c\left[(C_3-C_9/2)\mathcal{M}_{e,\pi K^*} +\left(C_5-C_7/2\right)\mathcal{M}_{e,\pi K^*}^R-\left(3C_{10}/2\right)\mathcal{M}_{e,K^*\pi}-\left(3C_8/2\right)\mathcal{M}_{e,K^*\pi}^P \right. \nonumber \\
		&\left.\left.+(C_3-C_9/2)\mathcal{M}_{a,K^*\pi}+\left(C_5-C_7/2\right)\mathcal{M}_{a,K^*\pi}^R\right]\right\}, 
\end{align}
\begin{align} 
	\mathcal{M}&(B^-\to \bar{K}^0\rho^-) 
		=V_u\left[a_1F_{a,K\rho}+\left(C_1/N_c\right)\mathcal{M}_{a,K\rho}\right]-V_t\left\{\left(a_4-a_{10}/2\right)F_{e,\rho K}+\left(a_6-a_8/2\right)F_{e,\rho K}^P \right. \nonumber \\ 
		&+\left(a_4+a_{10}\right)F_{a,K\rho} +\left(a_6+a_8\right)F_{a,K\rho}^P+{1}/{N_c}\left[\left(C_3-C_9/2\right)\mathcal{M}_{e,\rho K}+\left(C_5-C_7/2\right)\mathcal{M}_{e,\rho K}^R \right.\nonumber \\
		&\left.\left. +(C_3+C_9)\mathcal{M}_{a,K\rho}+\left(C_5+C_7\right)\mathcal{M}_{a,K\rho}^R\right]\right\}, \\
	\sqrt{2}\mathcal{M}&(B^-\to K^-\rho^0)
		=V_u\left[a_1F_{e,\rho K}+a_2F_{e,K\rho}+a_1F_{a,K\rho}+1/N_c\left(C_1\mathcal{M}_{e,\rho K}+C_2\mathcal{M}_{e,K\rho}+C_1\mathcal{M}_{a,K\rho}\right)\right] \nonumber \\
		&-V_t\left\{\left(a_4+a_{10}\right)F_{e,\rho K} +\left(a_6+a_8\right)F_{e,\rho K}^P
			+\left(3a_9/2\right)F_{e,K\rho}+\left(3a_7/2\right)F_{e,K\rho}^R +\left(a_4+a_{10}\right)F_{a,K\rho}\right. \nonumber \\ 
		&+\left(a_6+a_8\right)F_{a,K\rho}^P+1/N_c\left[(C_3+C_9)\mathcal{M}_{e,\rho K}+(C_5+C_7)\mathcal{M}_{e,\rho K}^R
			+\left(3C_{10}/2\right)\mathcal{M}_{e,K\rho}+\left(3C_8/2\right)\mathcal{M}_{e,K\rho}^P \right. \nonumber \\
		&\left.+(C_3+C_9)\mathcal{M}_{a,K\rho}+(C_5+C_7)\mathcal{M}_{a,K\rho}^R\right\}, \\
	\mathcal{M}&(\bar{B}^0\to K^-\rho^+) 
		=V_u\left[a_1F_{e,\rho K}+\left(C_1/N_c\right)\mathcal{M}_{e,\rho K}\right]
			-V_t\left\{\left(a_4+a_{10}\right)F_{e,\rho K}+\left(a_6+a_8\right)F_{e,\rho K}^P \right. \nonumber\\ 
		&+\left(a_4-a_{10}/2\right)F_{a,K\rho}+\left(a_6-a_8/2\right)F_{a,K\rho}^P
			+1/N_c\left[(C_3+C_9)\mathcal{M}_{e,\rho K}+(C_5+C_7)\mathcal{M}_{e,\rho K}^R \right. \nonumber \\ 
		&\left.\left.+(C_3-C_9/2)\mathcal{M}_{a,K\rho}+(C_5-C_7/2)\mathcal{M}_{a,K\rho}^R\right]\right\}, \\
	-\sqrt{2}\mathcal{M}&(\bar{B}^0\to \bar{K}^0\rho^0) 
		=V_u\left[-a_2F_{e,K\rho}-\left(C_2/N_c\right)\mathcal{M}_{e,K\rho}\right] 
			-V_t\left\{\left(a_4-a_{10}/2\right)F_{e,\rho K}+\left(a_6-a_8/2\right)F_{e,\rho K}^P \right. \nonumber \\
		&-\left(3a_9/2\right)F_{e,K\rho}-\left(3a_7/2\right)F_{e,K\rho}^R +\left(a_4-a_{10}/2\right)F_{a,K\rho} +\left(a_6-a_8/2\right)F_{a,K\rho}^P \nonumber \\
		&+ 1/N_c\left[(C_3-C_9/2)\mathcal{M}_{e,\rho K} +\left(C_5-C_7/2\right)\mathcal{M}_{e,\rho K}^R-\left(3C_{10}/2\right)\mathcal{M}_{e,K\rho}-\left(3C_8/2\right)\mathcal{M}_{e,K\rho}^P \right. \nonumber \\
		&\left.\left.+(C_3-C_9/2)\mathcal{M}_{a,K\rho}+\left(C_5-C_7/2\right)\mathcal{M}_{a,K\rho}^R\right]\right\}, \label{eq:KRhoamp}
\end{align}
\end{widetext}

With the amplitude for each decay channel given in Eqs.~\eqref{eq:KstarPiamp}-\eqref{eq:KRhoamp}, 
we can calculate the branching ratio and direct \textit{CP} violation, $\text{Br}$ and $A_{CP}$, by 
\begin{align}
	\Gamma(B\to M_1M_2)&=\dfrac{G_F^2m_B^3}{128\pi}\left|\mathcal{M}(B\to M_1M_2)\right|^2, 
\end{align}
\begin{align}
	\text{Br}(B\to M_1M_2)&=\Gamma(B\to M_1M_2)/\Gamma_B, 
\end{align}
\begin{align}
	A_{CP}&= \dfrac{\left(\Gamma(\bar{B}\to \bar{M_1}\bar{M_2}) - \Gamma(B\to M_1M_2)\right)}{\left(\Gamma(\bar{B}\to \bar{M_1}\bar{M_2}) + \Gamma(B\to M_1M_2)\right)}, 
\end{align}
where $\Gamma_B$ is the total decay width of $B$ meson. 

\subsection{The next-to-leading order corrections \label{subsec:nloamp}}
In the PQCD approach, there are several important contributions should be taken into account to improve the accuracy of the results for $B\to K^*\pi$ and $B\to K\rho$ decays. 
At first, it is necessary to determine the Wilson coefficients $C_i$ and the strong running coupling $\alpha_s$ at the next-to-leading-order (NLO) in QCD. 
Besides the diagrams shown in Fig.~\ref{fig:eightdiagram}, there are also three types of Feynman diagrams which contribute to the results significantly: 
the vertex correction, the quark loop correction, and the chromomagnetic penguin correction \cite{Li:2005kt}. 

The vertex corrections can be absorbed into the Wilson coefficients with the notation defined in Eq.~\eqref{eq:Wilsona}
\begin{align}
	a_{1,2}(\mu)&\to a_{1,2}(\mu)+\dfrac{\alpha_s(\mu)}{4\pi}C_F\dfrac{C_{1,2}(\mu)}{N_c}V_{1,2}(M), \nonumber\\
    a_{i,j}(\mu)&\to a_{i,j}(\mu)+\dfrac{\alpha_s(\mu)}{4\pi}C_F\dfrac{C_{i+1,j-1}(\mu)}{N_c}V_{i,j}(M), 
\end{align}
with $i=3,5,7,9$ and $j=4,6,8,10$. 
The meson $M$ is the meson that emitted from the weak vertex in Fig.~\ref{fig:eightdiagram}(a) and (b).  
The vertex function for the pseudoscalar meson, $V_i(M)$, in the naive dimensional regularization scheme are given by \cite{Beneke:1999br, Beneke:2000ry, Beneke:2001ev}
\begin{align} 
	V_i(M) =& 12\ln(m_b/\mu) - 18 + \dfrac{2\sqrt{2N_c}}{f_M}\int_0^1dx\phi_M^A(x)g(x), \nonumber \\
		&\hspace{8em} \text{for } i = 1,2,3,4,9,10  \nonumber \\ 
	V_i(M) =& -12\ln(m_b/\mu) + 6 -\dfrac{2\sqrt{2N_c}}{f_M}\int_0^1dx\phi_M^A(x)g(\bar{x}), \nonumber \\
		&\hspace{12em} \text{for } i = 5,7 \nonumber \\  
	V_i(M) =& \dfrac{2\sqrt{2N_c}}{f_M}\int_0^1dx\phi_M^P(x)\left[h(x)-6\right], \quad \text{for } i=6,8 \label{eq:vcv}
\end{align}
where $f_M$ is the decay constant, $\phi_M^{A,P}$ are the LCDAs, and $\bar{x}=1-x$. 
The hard kernels, $g(x)$ and $h(x)$, are 
\begin{align}
	g(x)=&3\left(\dfrac{1-2x}{1-x}\ln x-i\pi\right)+\biggl[2\text{Li}_2(x)-\ln^2x \nonumber \\
		&+\dfrac{2\ln x}{1-x}-(3+2i\pi)\ln x-(x\leftrightarrow 1-x)\biggr], \\
	h(x)=&2\text{Li}_2(x)-\ln^2x-(1+2i\pi)\ln x-(x\leftrightarrow 1-x).
\end{align}
For the vector meson, the vertex function can be derived from the Eq.~\eqref{eq:vcv} by making the substitution: $\phi_M^{A,P}\to \phi_V, \phi_V^s$.

Given the similar topological structure of the quark loop diagrams with the penguin diagrams, 
the effects of the quark loop and chromomagnetic penguin corrections can also be integrated into the Wilson coefficients respectively as \cite{Li:2005kt}
\begin{align}
	a_{4,6}(\mu)&\to a_{4,6}(\mu)+\dfrac{\alpha_s(\mu)}{9\pi}\sum\limits_{q=u,c,t}\frac{V_{qb}V_{qs}^*}{V_{tb}V_{ts}^*}C^{(q)}(\mu,\langle l^2\rangle), \\
	a_{4,6}(\mu)&\to a_{4,6}(\mu)-\dfrac{\alpha_s(\mu)}{9\pi}\dfrac{2m_B}{\sqrt{\langle l^2\rangle}}C_{8g}^\text{eff}(\mu),
\end{align}
where $\langle l^2\rangle$ represents the mean squared momentum for the virtual gluon in the decay diagrams, 
and the effective Wilson coefficient $C_{8g}^\text{eff}$ is defined by $C_{8g}^\text{eff}=C_{8g}+C_5$ \cite{Buchalla:1995vs}. 
The function $C^{(q)}(\mu, l^2)$ are 
\begin{align}
	C^{(q)}(\mu,l^2)&=\left[G^{(q)}(\mu,l^2)-2/3\right]C_2(\mu), \nonumber \\ 
		&\hspace{10em} \text{for } q=u,c \nonumber \\
    C^{(q)}(\mu,l^2)&=\left[G^{(s)}(\mu,l^2)-2/3\right]C_3(\mu) \nonumber \\
		&+\sum\limits_{q^{\prime\prime}=u,d,s,c}G^{(q^{\prime\prime})}(\mu,l^2)\left[C_4(\mu)+C_6(\mu)\right], \nonumber \\
		&\hspace{10em} \text{for } q=t,
\end{align}
with the $G^{(q)}(\mu,l^2)$ given by 
\begin{align}
	&G^{(q)}=-4\int_0^1dxx(1-x)\ln\frac{m_q^2-x(1-x)l^2-i\epsilon}{\mu^2}. 
\end{align}
We neglect the quark loop corrections induced by the electroweak operators here. 

\section{The contributions from soft form factors and color-octet matrix elements \label{sec:soft}}
With the results of Feynman diagrams shown in Fig.~\ref{fig:eightdiagram}, 
we can check the reliability of the perturbative calculation by introducing an infrared cutoff energy scale in $\mu_c$, 
where contributions with scale $\mu>\mu_c$ are viewed as perturbative contributions and $\mu<\mu_c$ nonperturbative contribution. 
One can adjust the value of scale $\mu_c$ gradually around 1 GeV. 
As discussed in Refs.~\cite{Lu:2022hbp, Wang:2022ihx}, the soft part in the region where $\alpha_s/\pi > 0.2$ contributes insignificantly for the nonfactorizable diagrams. 
However, in the factorizable diagrams illustrated in Fig.~\ref{fig:eightdiagram}(a)-(b) and Fig.~\ref{fig:eightdiagram}(g)-(h), 
the soft part accounts for more than $40\%$ of the contributions. 
To improve the applicability of the perturbative calculation, the soft contributions with the scale $\mu<\mu_c=1~\text{GeV}$ are attributed to a few soft form factors. 
Here, we can introduce two kinds of soft form factors to absorb the soft contributions in Fig.~\ref{fig:eightdiagram}(a)-(b) and Fig.~\ref{fig:eightdiagram}(g)-(h) respectively: 
the soft transition form factor and the soft production form factor. 
We also incorporate the effects of the color-octet operators, which have often been disregarded in general, and introduce the color-octet parameters as in Refs.~\cite{Wang:2022ihx, Lu:2024jzn}. 

\subsection{Soft transition form factors \label{subsec:softbm}}
In the calculations of the $B\to K^*\pi$ and $B\to K\rho$ decays, transition form factors for $B\to P$ and $B\to V$ are involved, 
where $P$ and $V$ denotes pseudoscalar meson and vector meson respectively. 
The form factors are defined by 
{
\allowdisplaybreaks 
\setlength{\jot}{10pt} 
\begin{align} \label{eq:BMff}
	\langle P|&\bar{q}\gamma_\mu(1-\gamma_5)b|\bar{B}\rangle \nonumber \\
		=& \left(p_B+p_P-\dfrac{m_B^2-m_P^2}{q^2}q\right)_\mu F_+(q^2) \nonumber \\
		&+ \dfrac{m_B^2-m_P^2}{q^2}q_\mu F_0(q^2), \\
	\langle V|&\bar{q}\gamma_\mu(1-\gamma_5)b|\bar{B}\rangle \nonumber \\
		=&\dfrac{2\epsilon_{\mu\nu\rho\sigma}}{m_B+m_V}\epsilon_V^\nu p_B^\rho p_V^\sigma V(q^2) +i\dfrac{\epsilon_V\cdot q}{q^2}2m_Vq_\mu A_0(q^2) \nonumber \\ 
		&+i\left[\epsilon_{V\mu}(m_B+m_V)A_1(q^2) - \dfrac{\epsilon_V\cdot q}{q^2}2m_Vq_\mu A_3(q^2) \right. \nonumber \\
		&-\left.\dfrac{\epsilon_V\cdot q}{m_B+m_V}\left(p_B+p_V\right)_\mu A_2(q^2)\right], 
\end{align}
}
where $\epsilon_V$ is the polarization vector and $q=p_B-p_{P(V)}$. 
Considering the soft part, we can express the involved transition form factor as 
\begin{align} \label{eq:BMff2}
	F_+^{BP} &= h_+^{BP} + \xi^{BP},\quad A_0^{BV} = h_{A0}^{BV} + \xi_{A0}^{BV}, 
\end{align}
where the $h$'s represent the hard contributions computed by PQCD approach \cite{Lu:2021gwt} and the $\xi$'s are the soft form factors which serve to absorb the nonperturbative effects. 

\subsection{Soft production form factors \label{subsec:softmm}}
The production form factor is defined by the final-state $M_1M_2$ to vacuum matrix element with the insertion of scalar and pseudoscalar four-quark operators: 
\begin{align} \label{eq:M1M2ff}
	\langle M_1M_2 |(S+P)|0 \rangle = -\dfrac{1}{2}\sqrt{\mu_1\mu_2}F^{M_1M_2},
\end{align}
where $\mu_{1,2}$ are chiral masses as mentioned after Eq. \eqref{k-wave}.
Similarly, the production form factor can also be separated into hard and soft parts by the critical scale $\mu_c$
\begin{align} \label{eq:M1M2ff2}
	F^{M_1M_2} &= h^{M_1M_2} + \xi^{M_1M_2}.
\end{align}
In this context, the soft production form factor $\xi^{M_1M_2}$ absorbs the contributions from the nonperturbative region of $\mu<\mu_c$ in Fig.~\ref{fig:eightdiagram}(g) and (h). 

The decay amplitudes listed in Eqs.~\eqref{eq:KstarPiamp}-\eqref{eq:KRhoamp} are modified by the inclusion of $\xi^{BM}$ and $\xi^{M_1M_2}$ 
combined with the corresponding Wilson coefficients and CKM matrix elements \cite{Wang:2022ihx, Lu:2024jzn}. 

\subsection{Color-octet parameters \label{subsec:softco}}
In this work, we take into account the contributions from the situation that the quark-antiquark pairs in the final-state are in color-octet states after the hard short-distance interactions in QCD. 
Since the final-state mesons should be in color-singlet states, the color-octet contributions are usually been disregarded in general. 
However, the quark-antiquark pairs in color-octet states can transform into color-singlet states by exchanging soft gluons at the distance of the hadronic scale. 
The color-octet contributions play an important role in our analysis about $B\to\pi\pi$, $\pi K$ and other $PP$ channels \cite{Lu:2022hbp,Wang:2022ihx,Lu:2024jzn}. 
As stated above, the transition of quark-antiquark pairs from color-octet to color-singlet state is dominated by long-distance dynamics, 
rather than short-distance interaction associated with hard gluon exchange. 

To extract the color-octet contribution, we need to consider the case that the final quark-antiquark pairs in $B$ decays shown in Fig.~\ref{fig:eightdiagram} are in color non-singlet state, 
and analyze the color factors in each diagram in Fig.~\ref{fig:eightdiagram}. 
With the relation of color $\text{SU(3)}_c$ generators, $T^a_{ik}T^a_{jl}=-\delta_{ik}\delta_{jl}/{2N_c}+\delta_{il}\delta_{jk}/2$, 
the quark-antiquark pairs can be decomposed into color-singlet and color-octet parts.
Assuming that the distribution amplitudes for quark-antiquark pairs in color-singlet and color-octet states are the same, 
the color-octet results for each diagram differ from those of the color-singlet case, listed in Eqs.~\eqref{eq:fe}-\eqref{eq:map}, solely by a color factor. 
Since a thorough discussion about the color factor has been provided in Refs.~\cite{Wang:2022ihx, Lu:2024jzn}, we present the color-octet results directly here: 
\begin{align} \label{eq:fm8}
	F_{e,M_1M_2}^{(P,R),8} &\equiv 2N_c^2 F_{e,M_1M_2}^{(P,R),a} - \dfrac{N_c}{C_F} F_{e,M_1M_2}^{(P,R),b}, \nonumber\\
    \mathcal{M}_{e,M_1M_2}^{(P,R),8} &\equiv 2 N_c^2 \mathcal{M}_{e,M_1M_2}^{(P,R),c} - \dfrac{N_c}{C_F} \mathcal{M}_{e,M_1M_2}^{(P,R),d}, \nonumber\\
    \mathcal{M}_{e,M_1M_2}^{(P,R),8^\prime} &\equiv \dfrac{N_c^2}{C_F} \mathcal{M}_{e,M_1M_2}^{(P,R)}, \  
	\mathcal{M}_{a,M_1M_2}^{(P,R),8} \equiv -\dfrac{N_c}{C_F} \mathcal{M}_{a,M_1M_2}^{(P,R)},\nonumber\\
	F_{a,M_1M_2}^{(P,R),8} &\equiv -\dfrac{N_c^2}{C_F} F_{a,M_1M_2}^{(P,R)},
\end{align}
where the superscripts $8$ and $8^\prime$ denote the color-octet result, and $a,b,c,d$ indicate the contributions of the corresponding diagrams shown in Fig.~\ref{fig:eightdiagram}. 

Since the final-state pseudoscalar and vector mesons must be in color-singlet states, the color-octet quark-antiquark pairs should be transformed into color-singlet states by exchanging soft gluons. 
The transition of quark-antiquark pairs from the color-octet to the color-singlet state occurs in the soft region, where reliable calculation cannot be achieved in the perturbative theory. 
To account for this effect, we introduce the multiplicative parameters $Y_{F,M}^8$. 
By incorporating color-octet diagram results with parameters $Y_{F,M}^8$ and some constants, the color-octet contribution can be described completely. 
The amplitudes for the $B\to K^*\pi$ decays are modified as 
\begin{widetext}
\allowdisplaybreaks 
\setlength{\jot}{10pt} 
\begin{align}
	\mathcal{M}(B^-&\to \bar{K}^{*0}\pi^-)\to \mathcal{M}(B^-\to \bar{K}^{*0}\pi^-)
			+\left\{V_ua_1F_{a,K^*\pi}^8 - V_t\left[\left(C_3-C_9/2\right)F_{e,\pi K^*}^8 + \left(C_5-C_7/2\right)F_{e,\pi K^*}^{P,8} \right.\right.\nonumber\\
		&\left.\left.+\left(a_4+a_{10}\right)F_{a,K^*\pi}^8 + \left(a_6+a_8\right)F_{a,K^*\pi}^{P,8}\right]\right\}Y_F^8 
			+\left\{V_uC_1\mathcal{M}_{a,K^*\pi}^8 - V_t\Bigl[(C_3-C_9/2)\mathcal{M}_{e,\pi K^*}^8 \right. \nonumber \\ 
		&+(C_4-C_{10}/2)\mathcal{M}_{e,\pi K^*}^{8^\prime} + (C_5-C_7/2)\mathcal{M}_{e,\pi K^*}^{R,8} + (C_6-C_8/2)\mathcal{M}_{e,\pi K^*}^{R,8^\prime} 
			+(C_3+C_9)\mathcal{M}_{a,K^*\pi}^8 \nonumber\\
		&\left. +(C_5+C_7)\mathcal{M}_{a,K^*\pi}^{R,8} \Bigr]\right\} Y_M^8, \label{eq:KstarPiamp18}\\
	\sqrt{2}\mathcal{M}(B^-&\to K^{*-}\pi^0)\to \sqrt{2}\mathcal{M}(B^-\to K^{*-}\pi^0) 
			+\Bigl\{V_u\left(C_1F_{e,\pi K^*}^8 + C_2F_{e,K^*\pi}^8 + a_1F_{a,K^*\pi}^8\right) -V_t\Bigl[(C_3 + C_9)F_{e,\pi K^*}^8 \nonumber \\ 
		&+(C_5 + C_7)F_{e,\pi K^*}^{P,8} + \left(3C_{10}/2\right)F_{e,K^*\pi}^8 + \left(3C_8/2\right)F_{e,K^*\pi}^{R,8} 
			+ (a_4+a_{10})F_{a,K^*\pi}^8 + (a_6+a_8)F_{a,K^*\pi}^{P,8}\Bigr]\Bigr\} Y_{F}^{8} \nonumber\\
		&+\left\{V_u\left(C_1\mathcal{M}_{e,\pi K^*}^8 + C_2\mathcal{M}_{e,\pi K^*}^{8^\prime} + C_2\mathcal{M}_{e,K^*\pi}^8 
			+ C_1\mathcal{M}_{e,K^*\pi}^{8^\prime} + C_1\mathcal{M}_{a,K^*\pi}^8\right) 
			-V_t\Bigl[(C_3 + C_9)\mathcal{M}_{e,\pi K^*}^8 \right.\nonumber \\ 
		&+(C_4+C_{10})\mathcal{M}_{e,\pi K^*}^{8^\prime} + (C_5 + C_7)\mathcal{M}_{e,\pi K^*}^{R,8} + (C_6 + C_8)\mathcal{M}_{e,\pi K^*}^{R,8^\prime}
			+\left(3C_{10}/2\right)\mathcal{M}_{e,K^*\pi}^8 + \left(3C_9/2\right)\mathcal{M}_{e,K^*\pi}^{8^\prime} \nonumber \\
		&\left.+\left(3C_8/2\right)\mathcal{M}_{e,K^*\pi}^{P,8} + \left(3C_7/2\right)\mathcal{M}_{e,K^*\pi}^{P,8^\prime} + (C_3 + C_9)\mathcal{M}_{a,K^*\pi}^8 
			+(C_5 + C_7)\mathcal{M}_{a,K^*\pi}^{R,8}\Bigr]\right\} Y_M^8, \label{eq:KstarPiamp28}\\
	\mathcal{M}(\bar{B}^0&\to K^{*-}\pi^+) \to \mathcal{M}(\bar{B}^0\to K^{*-}\pi^+) 
		+\left\{V_uC_1F_{e,\pi K^*}^8 -V_t\left[(C_3 + C_9)F_{e,\pi K^*}^8 + (C_5 + C_7)F_{e,\pi K^*}^{P,8} \right.\right.\nonumber\\
		&\left.\left.+(a_4-a_{10}/2)F_{a, K^*\pi}^8 + (a_6-a_8/2)F_{a,K^*\pi}^{P,8}\right]\right\} Y_F^8 
			+\left\{V_u\left(C_1\mathcal{M}_{e,\pi K^*}^8 + C_2\mathcal{M}_{e,\pi K^*}^{8^\prime} \right)\right. \nonumber \\ 
		&-V_t\Bigl[(C_3 + C_9)\mathcal{M}_{e,\pi K^*}^8 + (C_4 + C_{10})\mathcal{M}_{e,\pi K^*}^{8^\prime} 
			+ (C_5 + C_7)\mathcal{M}_{e,\pi K^*}^{R,8} + (C_6 + C_8)\mathcal{M}_{e,\pi K^*}^{R,8^\prime} \nonumber \\
		&\left.+(C_3 - C_9/2)\mathcal{M}_{a,K^*\pi}^8 + (C_5 - C_7/2)\mathcal{M}_{a,K^*\pi}^{R,8}\Bigr]\right\} Y_M^8, \label{eq:KstarPiamp38}\\
	-\sqrt{2}\mathcal{M}(\bar{B}^0&\to \bar{K}^{*0}\pi^0) \to -\sqrt{2}\mathcal{M}(\bar{B}^0\to \bar{K}^{*0}\pi^0) 
			+\left\{-V_uC_2F_{e,K^*\pi}^8 -V_t\left[\left(C_3 - C_9/2\right)F_{e,\pi K^*}^8 + \left(C_5 - C_7/2\right)F_{e,\pi K^*}^{P,8} \right.\right.\nonumber \\ 
		&\left.\left. -\left(3C_{10}/2\right)F_{e,K^*\pi}^8 - \left(3C_8/2\right)F_{e,K^*\pi}^{R,8} + (a_4-a_{10}/2)F_{a,K^*\pi}^8 
			+ (a_6-a_8/2)F_{a,K^*\pi}^{P,8} \right]\right\}Y_F^8  \nonumber \\
		&+\left\{-V_u\left(C_2\mathcal{M}_{e,K^*\pi}^8 + C_1\mathcal{M}_{e,K^*\pi}^{8^\prime}\right) 
			-V_t\left[(C_3 - C_9/2)\mathcal{M}_{e,\pi K^*}^8 + (C_4 - C_{10}/2)\mathcal{M}_{e,\pi K^*}^{8^\prime} \right.\right.\nonumber \\
		&+(C_5 - C_7/2)\mathcal{M}_{e,\pi K^*}^{R,8} + (C_6 - C_8/2)\mathcal{M}_{e,\pi K^*}^{R,8^\prime} - \left(3C_{10}/2\right)\mathcal{M}_{e,K^*\pi}^8 
			- \left(3C_9/2\right)\mathcal{M}_{e,K^*\pi}^{8^\prime} \nonumber \\ 
		&\left.\left. -\left(3C_8/2\right)\mathcal{M}_{e,K^*\pi}^{P,8} - \left(3C_7/2\right)\mathcal{M}_{e,K^*\pi}^{P,8^\prime} + (C_3 - C_9/2)\mathcal{M}_{a,K^*\pi}^8 
			+ (C_5 - C_7/2)\mathcal{M}_{a,K^*\pi}^{R,8} \right]\right\}Y_M^8. \label{eq:KstarPiamp48}
\end{align}
\end{widetext}
For the $B\to K\rho$ decays, the amplitudes with the color-octet contribution included can be obtained from Eqs.~\eqref{eq:KstarPiamp18}-\eqref{eq:KstarPiamp48} through 
the substitution of the mesons: $K^*\to K$, $\pi\to\rho$. 

\section{Numerical analysis and discussion \label{sec:num}}
In the numerical calculation, the input parameters include the parameters for the $B$ meson and light meson distribution amplitudes which are shown in Eqs.~\eqref{eq:a1234} and ~\eqref{eq:lcdapara}. 
Other parameters are the $W$ boson mass, 
as well as the masses, decay constants, and lifetimes of various mesons. 
These parameters are presented as follows (masses and decay constants are expressed in the units of $\text{GeV}$), which are mainly quoted from the PDG \cite{ParticleDataGroup:2024cfk} 
{
\allowdisplaybreaks 
\begin{align}
	m_W &= 80.4,	\ m_B = 5.279,	\ \tau_{B^\pm/B^0} = 1.638/1.519~\text{ps}, \nonumber \\
	m_\pi &= 0.14,	\ m_K = 0.494,	\ m_\rho = 0.775,	\ m_{K^*} = 0.892, \nonumber \\
	r_\pi &= 0.33,	\ r_K = 0.32,	\ r_\rho = 0.15,	\ r_{K^*} = 0.17, \nonumber \\
	f_B &= 0.21, 	\ f_\pi = 0.13, \ f_K = 0.16,		\ f_\rho^\parallel = 0.216, \nonumber \\
	f_\rho^\perp &= 0.165,\ f_{K^*}^\parallel = 0.220,	\ f_{K^*}^\perp = 0.185. 
\end{align}
}

For nonperturbative parameters, the soft transition form factors, the soft production form factors, and the color-octet parameters, their values are determined through different methods. 

The $B\to P$ and $B\to V$ transition form factors have been extensively studied in both theoretical and experimental aspects \cite{Parrott:2022rgu, Bailey:2015dka, Ball:2004ye, Ball:2004rg, Khodjamirian:2006st, Bharucha:2015bzk, Gubernari:2018wyi, Gao:2019lta, Ivanov:2007cw, Ivanov:2011aa, Belle-II:2018jsg}. 
Given that the $B$ meson is much heavier than the final-state pseudoscalar and vector mesons concerned in this work, it is sufficient to take the value of the transition form factor 
at the large recoil point $q^2=0$. The transition form factors are taken as \cite{Parrott:2022rgu, Bharucha:2015bzk,Gao:2019lta, Ivanov:2011aa}
\begin{align}
	F_+^{B\pi} &= 0.27\pm 0.02, &F_+^{BK} &= 0.33\pm 0.04, \nonumber \\
	A_0^{B\rho} &= 0.32\pm 0.05, &A_0^{BK^*} &= 0.36\pm 0.04.
\end{align}
With the cutoff scale taken $\mu_c=1~\text{GeV}$, we can calculate the hard contributions to $F_+^{BP}$ and $A_0^{BV}$ in the framework of PQCD. 
The values of $h_+^{BP}$ and $h_{A0}^{BV}$ are given by
\begin{align}
	h_+^{B\pi} &= 0.23\pm 0.01, &h_+^{BK} &= 0.29\pm 0.02, \nonumber \\
	h_{A0}^{B\rho} &= 0.18\pm 0.01, &h_{A0}^{BK^*} &= 0.20\pm 0.01, 
\end{align}
where the uncertainties arise from the meson LCDA parameters. 
The soft transition form factors can be derived from the Eq.~\eqref{eq:BMff2} as 
\begin{align}
	\xi_+^{B\pi} &= 0.04\pm 0.01, &\xi_+^{BK} &= 0.04\pm 0.02, \nonumber \\
	\xi_{A0}^{B\rho} &= 0.14\pm 0.04, &\xi_{A0}^{BK^*} &= 0.16\pm 0.03. 
\end{align}

Unlike the transition form factor, which can be extracted from the semileptonic decay, the production form factor cannot be derived from the experimental result directly. 
Furthermore, there is no reliable way to calculate the color-octet parameter due to the intricate nonperturbative dynamics. 
We have to constrain these parameters by experimental results, i.e. the branching ratios and direct $CP$ violations of $B\to K^*\pi$ and $B\to K\rho$ decays. 
With the global numerical fitting, these parameters which can explain all the experimental data involved in this work are shown below: 
{
\allowdisplaybreaks 
\begin{align}
	\mu_{K^*\pi}F^{K^*\pi} &= 0.143^{+0.031}_{-0.028}\exp\left[\left(-0.676^{+0.067}_{-0.068}\right)\pi i\right], \nonumber \\
	Y_{F,K^*\pi}^8 &= 0.208^{+0.017}_{-0.018}\exp\left[\left( 0.894^{+0.033}_{-0.032}\right)\pi i\right], \nonumber \\
	Y_{M,K^*\pi}^8 &= 0.158^{+0.033}_{-0.030}\exp\left[\left(-0.922^{+0.063}_{-0.061}\right)\pi i\right], \\
	\mu_{K\rho}F^{K\rho} &= 0.164^{+0.043}_{-0.040}\exp\left[\left(-0.253^{+0.134}_{-0.121}\right)\pi i\right], \nonumber \\
	Y_{F,K\rho}^8 &= 0.194^{+0.026}_{-0.026}\exp\left[\left(-0.724^{+0.190}_{-0.213}\right)\pi i\right], \nonumber \\ 
	Y_{M,K\rho}^8 &= 0.201^{+0.024}_{-0.024}\exp\left[\left(-0.571^{+0.139}_{-0.134}\right)\pi i\right],
\end{align}
}
where $\mu_{M_1M_2}=\sqrt{\mu_{M_1}\mu_{M_2}}$. 

The numerical results of branching ratios and $CP$ violations for eight decay channels are shown in Table~\ref{tab:bracpresults}. 
The column denoted by ``$\text{LO}_{\text{NLOWC}}$'' shows the results calculated from LO diagrams in QCD and with NLO Wilson coefficients. 
The ``NLO'' column is for the theoretical results including NLO corrections. 
The column with ``NLO+$\xi^{BM},\xi^{M_1M_2}$" is for the results up to NLO corrections in QCD plus the contributions of soft transition and production form factors. 
In the ``NLO+Soft'' column, soft contributions - including the soft transition form factor, soft production form factor, and color-octet contributions 
- are taken into account, which are the total final results of this work. 
The experimental data taken from the PDG \cite{ParticleDataGroup:2024cfk} are shown in last column for comparison. 
\begin{table*}
	\caption{\label{tab:bracpresults} Branching ratios $(\times 10^{-6})$ and \textit{CP} violations of the $B\to K^*\pi$ and $B\to K\rho$ decays.}
	\renewcommand\arraystretch{1.2}
	\begin{ruledtabular}
		\begin{tabular}{ldddcc}
		Mode &\multicolumn{1}{c}{$\text{LO}_{\text{NLOWC}}$} & \multicolumn{1}{c}{NLO} &\multicolumn{1}{c}{NLO+$\xi^{BM},\xi^{M_1M_2}$} & NLO+Soft & Data \cite{ParticleDataGroup:2024cfk} \\ 
		\hline
		Br$(B^+\to K^{*0}\pi^+)$      &5.0   &5.0   &16.8  &$ 9.6 ^{+0.9+0.2+0.4}_{-0.9-0.4-0.4}$ 				&$10.1\pm 0.8$ \\          
		Br$(B^+\to K^{*+}\pi^0)$      &4.1   &3.4   &9.0   &$ 5.9 ^{+0.5+0.2+0.2}_{-0.6-0.3-0.2}$ 				&$6.8 \pm 0.9$ \\          
		Br$(B^0\to K^{*+}\pi^-)$      &6.4   &5.4   &13.4  &$ 7.9 ^{+0.5+0.1+0.2}_{-0.5-0.2-0.2}$ 				&$7.5 \pm 0.4$ \\          
		Br$(B^0\to K^{*0}\pi^0)$      &1.8   &1.9   &6.1   &$ 2.9 ^{+0.3+0.0+0.1}_{-0.3-0.0-0.0}$ 				&$3.3 \pm 0.6$ \\          
		Br$(B^+\to K^0\rho^+)$        & 3.3  &4.2   &2.4   &$ 8.1 ^{+0.8+0.1+0.1}_{-0.7-0.1-0.1}$ 				&$7.3_{-1.2}^{+1.0}$ \\    
		Br$(B^+\to K^+\rho^0)$        & 1.3  &1.7   &0.9   &$ 3.2 ^{+0.6+0.1+0.1}_{-0.6-0.0-0.1}$ 				&$3.7 \pm 0.5$ \\          
		Br$(B^0\to K^+\rho^-)$        & 1.8  &2.2   &0.4   &$ 6.7 ^{+1.1+0.1+0.1}_{-1.0-0.1-0.1}$ 				&$7.0 \pm 0.9$ \\          
		Br$(B^0\to K^0\rho^0)$        & 1.8  &1.8   &0.8   &$ 4.3 ^{+0.5+0.0+0.1}_{-0.5-0.0-0.1}$ 				&$3.4 \pm 1.1$ \\ [0.5em]  
		$A_{CP}(B^+\to K^{*0}\pi^+)$  &-0.011& 0.008& 0.011&$ 0.001^{+0.002+0.000+0.002}_{-0.003-0.000-0.002}$ &$-0.021 \pm 0.032 $ \\     
		$A_{CP}(B^+\to K^{*+}\pi^0)$  &-0.39 &-0.15 & 0.10 &$-0.18 ^{+0.06 +0.00 +0.00 }_{-0.06 -0.00 -0.01 }$ &$-0.39 \pm 0.21 $ \\       
		$A_{CP}(B^0\to K^{*+}\pi^-)$  &-0.45 &-0.26 & 0.03 &$-0.27 ^{+0.05 +0.01 +0.03 }_{-0.06 -0.02 -0.02 }$ &$-0.27 \pm 0.04 $ \\       
		$A_{CP}(B^0\to K^{*0}\pi^0)$  & 0.04 &-0.04 &-0.07 &$-0.03 ^{+0.07 +0.02 +0.04 }_{-0.06 -0.03 -0.04 }$ &$-0.15 \pm 0.13$ \\        
		$A_{CP}(B^+\to K^0\rho^+)$    & 0.00 &0.00  &0.01  &$ 0.00 ^{+0.01 +0.00 +0.01 }_{-0.01 -0.00 -0.00 }$ &$-0.03 \pm 0.15 $ \\       
		$A_{CP}(B^+\to K^+\rho^0)$    & 0.459&0.423 &0.616 &$ 0.162^{+0.083+0.006+0.021}_{-0.078-0.009-0.021}$ &$ 0.160 \pm 0.021 $ \\     
		$A_{CP}(B^0\to K^+\rho^-)$    & 0.47 &0.44  &0.84  &$ 0.25 ^{+0.11 +0.00 +0.02 }_{-0.10 -0.01 -0.02 }$ &$ 0.20 \pm 0.11 $ \\       
		$A_{CP}(B^0\to K^0\rho^0)$    &-0.15 &-0.05 &-0.02 &$ 0.23 ^{+0.07 +0.01 +0.01 }_{-0.07 -0.00 -0.00 }$ &$ 0.04 \pm 0.20$ \\        
		\end{tabular}
	\end{ruledtabular}
\end{table*}

The comparison of the column ``NLO" and ``$\text{LO}_{\text{NLOWC}}$" in Table~\ref{tab:bracpresults} shows that the NLO contributions are generally small, 
which can only change the branching ratios slightly. 
Only for the decay modes where the LO contribution is suppressed by their color or isospin structures, 
the NLO contribution can change the branching ratios by about $20\%\sim 30\%$. 
This indicates that the introduction of the infrared cutoff scale $\mu_c$ makes the perturbation theory applicable. 
And it has been shown that varying the value of $\mu_c$ around $1~\text{GeV}$ will not change the physical results seriously \cite{Wang:2022ihx}, 
which indicates that taking $\mu_c=1~\mbox{GeV}$ is a reasonable choice. 
Columns ``NLO+$\xi^{BM},\xi^{M_1M_2}$" and ``NLO+Soft" imply that both the contributions of the soft factors and color-octet component are important to explain the experimental data. 
With the nonperturbative input parameters taking appropriate values, the theoretical results can be in good agreement with experimental data.

\section{Summary \label{sec:sum}} 
We revisit $B\to K^*\pi$ and $B\to K\rho$ decays with the modified PQCD approach, where an infrared cutoff scale $\mu_c$ is introduced. 
The contributions with scale $\mu>\mu_c$ are calculated with PQCD approach, while the dynamics below the infrared cutoff scale $\mu_c$ is viewed as nonperturbative interactions, 
which is treated in terms of soft form factors. 
The color-octet contributions are also introduced, which are essentially of long-distance nature. 
Our analysis implies that contributions of soft form factors and color-octet components are important for explaining the experimental data.

\begin{acknowledgments}
	This work is supported in part by the National Natural Science Foundation of China under Contracts No. 12275139, 11875168.
\end{acknowledgments}

\appendix{}
\section{Light meson distribution amplitudes \label{app:lightda}}
The light meson LCDA, $\phi(x,k_{q\perp})$, depends on not only the momentum fraction $x$ but also the transverse momentum $k_{q\perp}$. 
Here, we make an assumption that the $k_{q\perp}$ dependence of $\phi(x,k_{q\perp})$ is Gaussian distribution and can be separated from the $x$ part. 
With Fourier transformation, the LCDA in transverse coordinate $b$ space can be expressed as: 
\begin{equation}
	\phi(x,b)=\phi(x)\exp\left(-b^2/(4\beta^2)\right),
\end{equation}
where the oscillation parameter $\beta$ is taken to be $4.0~\textrm{GeV}^{-1}$, as discussed in Refs.~\cite{Wei:2002iu, Wang:2022ihx}. 

The LCDAs for pseudoscalar and vector meson have been studied systemically, 
see Refs.~\cite{Braun:1989iv, Ball:1996tb, Ball:1998je, Ball:2006wn, Ball:1998sk, Ball:2007rt, Zhong:2011rg, RQCD:2019osh, Cheng:2020vwr}.
In terms of the Gegenbauer polynomials $C_n^{m/2}$, the pseudoscalar meson LCDAs, $\phi_M^{A,P,T}$ are of the form
{
\allowdisplaybreaks 
\begin{widetext}
\begin{align} \label{eq:phiPAandphiPT}
	\phi_M^A(x)&=\dfrac{f_M}{2\sqrt{2N_c}}6x\bar{x}\left[1+a_{1M} C_1^{3/2}(t) + a_{2M} C_2^{3/2}(t)\right], \qquad 
	\phi_M^T(x) = \dfrac{f_M}{2\sqrt{2N_c}}\dfrac{1}{6}\dfrac{d}{dx}\phi_M^\sigma(x),
\end{align}
	\begin{align} \label{eq:phiPP}
	\phi_M^P(x)=&\dfrac{f_M}{2\sqrt{2N_c}}\Bigl\{1+3\rho_+^M(1+6a_{2M})-9\rho_-^M a_{1M} + \left[27\rho_+^M a_{1M}/2 - \rho_-^M\left(3/2 + 27a_{2M}\right)\right]C_1^{1/2}(t) \nonumber \\ 
		&+\left(30\eta_{3M}+15\rho_+^M a_{2M} - 3\rho_-^M a_{1M}\right) C_2^{1/2}(t) +\left(10\eta_{3M}\lambda_{3M} - 9\rho_-^M a_{2M}/2\right) C_3^{1/2}(t) 
			-3\eta_{3M}\omega_{3M} C_4^{1/2}(t) \nonumber \\ 
		&+3/2\left(\rho_+^M + \rho_-^M\right)\left(1-3a_{1M}+6a_{2M}\right)\ln x + 3/2\left(\rho_+^M - \rho_-^M\right)\left(1+3a_{1M}+6a_{2M}\right)\ln\bar{x}\Bigr\},
\end{align}
\begin{align}
	\phi_M^\sigma(x)=&6x\bar{x}\Bigl[1+3\rho_+^M/2 + 15\rho_+^M a_{2M} - 15\rho_-^M a_{1M}/2 + \left(3\rho_+^M a_{1M} - 15\rho_-^M a_{2M}/2\right) C_1^{3/2}(t) \nonumber \\ 
		&+\left(5\eta_{3M} - \eta_{3M}\omega_{3M}/2 + 3\rho_+^M a_{2M}/2\right) C_2^{3/2}(t) +\eta_{3M}\lambda_{3M} C_3^{3/2}(t)\Bigr] \nonumber \\ 
		&+9x\bar{x}\left(\rho_+^M + \rho_-^M\right)\left(1-3a_{1M}+6a_{2M}\right)\ln x +9x\bar{x}\left(\rho_+^M - \rho_-^M\right)\left(1+3a_{1M}+6a_{2M}\right)\ln\bar{x}, 
\end{align}
with $\bar{x} = 1-x$ and $t = 2x-1$. 
$\rho_{\pm}^M$ and $\eta_{3M}$ are defined by relevant decay constants and masses as: 
\begin{equation}
	\rho_+^M = \dfrac{(m_{q1}+m_{q2})^2}{m_M^2},\quad 	\rho_-^M = \dfrac{m_{q1}^2-m_{q2}^2}{m_M^2}, \quad 
	\eta_{3M} = \dfrac{f_{3M}}{f_M}\dfrac{m_{q1}+m_{q2}}{m_M^2}. 
\end{equation}
For the vector mesons, the LCDAs with the longitudinal polarization, $\phi_M$ and $\phi_M^{s, t}$, are given by: 
\begin{align} \label{eq:phiVandphiVs}
	\phi_M(x) &=\dfrac{f_M^\parallel}{2\sqrt{2N_c}} 6x\bar{x}\left[1+a_{1M}^\parallel C_1^{3/2}(t) + a_{2M}^\parallel C_2^{3/2}(t)\right], \qquad 
	\phi_M^s(x) = \dfrac{f_M^\perp}{2\sqrt{2N_c}} \dfrac{1}{2} \dfrac{d}{dx} \phi_M^u(x), 
\end{align}
\begin{align} \label{eq:phiVt}
	\phi_M^t(x) =& \dfrac{f_M^\perp}{2\sqrt{2N_c}}\Bigl\{3t C_1^{1/2}(t) + 3t a_{1M}^\perp C_2^{1/2}(t) 
			+ \left(3t a_{2M}^\perp + 15\kappa_{3M}^\perp - 3\lambda_{3M}^\perp/2\right) C_3^{1/2}(t) + 5\omega_{3M}^\perp C_4^{1/2}(t) \nonumber \\ 
		&+\dfrac{3}{2}\dfrac{m_{q1}+m_{q2}}{m_M}\dfrac{f_M^\parallel}{f_M^\perp}\left[1+8ta_{1M}^\parallel + 3(7-30x\bar{x})a_{2M}^\parallel 
			+t\left(1+3a_{1M}^\parallel+ 6a_{2M}^\parallel\right)\ln\bar{x} - t\left(1-3a_{1M}^\parallel+6a_{2M}^\parallel\right)\ln x\right] \nonumber \\ 
		&-\dfrac{3}{2}\dfrac{m_{q1}-m_{q2}}{m_M}\dfrac{f_M^\parallel}{f_M^\perp}t\left[2+9ta_{1M}^\parallel + 2(11- 30x\bar{x})a_{2M}^\parallel 
			+ \left(1+3a_{1M}^\parallel+6a_{2M}^\parallel\right)\ln\bar{x} +\left(1-3a_{1M}^\parallel+6a_{2M}^\parallel\right)\ln x \right]\Bigr\}
\end{align}
\begin{align} \label{eq:phiVu}
	\phi_M^u(x) =& 6x\bar{x}\left[1+\left(a_{1M}^\perp/3 + 5\kappa_{3M}^\perp/3\right) C_1^{3/2}(t) + \left(a_{2M}^\perp/6 + 5\omega_{3M}^\perp/18\right)C_{2}^{3/2}(t) 
			- \left(\lambda_{3M}^\perp/20\right) C_3^{3/2}(t)\right] \nonumber \\ 
		& +3\dfrac{m_{q1}+m_{q2}}{m_M}\dfrac{f_M^\parallel}{f_M^\perp}\left[x\bar{x}\left(1+2ta_{1M}^\parallel+3\left(7-5x\bar{x}\right)a_{2M}^\parallel\right) 
			+\left(1+3a_{1M}^\parallel+6a_{2M}^\parallel\right)\bar{x}\ln\bar{x} \right. \nonumber \\ 
		& \left. + \left(1-3a_{1M}^\parallel+6a_{2M}^\parallel\right) x\ln x \right]
			-3\dfrac{m_{q1}-m_{q2}}{m_M}\dfrac{f_M^\parallel}{f_M^\perp}\left[x\bar{x}\left(9a_{1M}^\parallel+10ta_{2M}^\parallel\right) 
			+\left(1+3a_{1M}^\parallel+6a_{2M}^\parallel\right)\bar{x}\ln\bar{x} \right. \nonumber \\ 
		& \left. -\left(1-3a_{1M}^\parallel+6a_{2M}^\parallel\right) x\ln x\right]. 
\end{align}
\end{widetext}
}
The hadronic parameters at the scale $\mu=1~\text{GeV}$ used in Eqs.~\eqref{eq:phiPAandphiPT}-\eqref{eq:phiVu} are listed below: 
\begin{subequations} \label{eq:lcdapara}
\allowdisplaybreaks 
\begin{align}
	a_{1\pi}&=0,\quad a_{2\pi}=0.25\pm 0.15,\quad f_{3\pi}=0.0045\pm 0.0015, \nonumber \\
	\omega_{3\pi}&=-1.5\pm 0.7,\quad \lambda_{3\pi}=0,
\end{align}
\begin{align}
	a_{1K}&=0.06\pm 0.03,\quad a_{2K}=0.25\pm 0.15,\nonumber \\
	f_{3K}&=0.0045\pm 0.0015, \quad \omega_{3K}=-1.2\pm 0.7,\nonumber \\ 
	\lambda_{3K}&=1.6\pm 0.4,
\end{align}
\begin{align}
	a_{1\rho}^\parallel &= a_{1\rho}^\perp = 0,\quad a_{2\rho}^\parallel = 0.15\pm 0.07, \quad a_{2\rho}^\perp = 0.14\pm 0.06, \nonumber \\
	\kappa_{3\rho}^\perp &= 0, \quad \omega_{3\rho}^\perp = 0.55\pm 0.25,\quad \lambda_{3\rho}^\perp = 0, 
\end{align}
\begin{align}
	a_{1K^*}^\parallel &= 0.03\pm 0.02,\quad a_{1K^*}^\perp = 0.04\pm 0.03, \nonumber \\ 
	a_{2K^*}^\parallel &= 0.11\pm 0.09, \quad a_{2K^*}^\perp = 0.10\pm 0.08, \nonumber \\
	\kappa_{3K^*}^\perp &= 0.003\pm 0.003,\quad \omega_{3K^*}^\perp = 0.3\pm 0.1, \nonumber \\ 
	\lambda_{3K^*}^\perp &= -0.025\pm 0.020.
\end{align}
\end{subequations}


\section{The formulas used in PQCD calculation \label{app:sudakovht}}
As mentioned above, the hard function $h$'s arise from the Fourier transformation of hard amplitudes. 
In Eqs.~\eqref{eq:fe}-\eqref{eq:map}, these $h$ functions are defined by 
{
\allowdisplaybreaks 
\setlength{\jot}{10pt} 
\begin{align}
	h_e&(x_1,x_2,b_1,b_2)=K_0(\sqrt{x_1x_2}m_Bb_1)[\theta(b_1-b_2) \nonumber \\ 
		&\times K_0(\sqrt{x_2}m_Bb_1)I_0(\sqrt{x_2}m_Bb_2) + (b_1 \leftrightarrow b_2)], \nonumber \\
	h_a&(x_1,x_2,b_1,b_2)=K_0(-i\sqrt{x_1x_2}m_Bb_1)[\theta(b_1-b_2) \nonumber \\ 
		&\times K_0(-i\sqrt{x_2}m_Bb_1)I_0(-i\sqrt{x_2}m_Bb_2)+ (b_1 \leftrightarrow b_2)], \nonumber \\
	h_{ne}&(x_1,x_2,x_3,b_1,b_2)=K_0(-i\sqrt{x_2x_3}m_Bb_2)[\theta(b_1-b_2) \nonumber \\ 
		&\times K_0(\sqrt{x_1x_3}m_Bb_1)I_0(\sqrt{x_1x_3}m_Bb_2)+ (b_1 \leftrightarrow b_2)], \nonumber \\	
	h_{na}^1&(x_1,x_2,b_1,b_2)=K_0(-i\sqrt{x_1x_2}m_Bb_1) \nonumber \\ 
		&\times[\theta(b_1-b_2) K_0(-i\sqrt{x_1x_2}m_Bb_1)I_0(-i\sqrt{x_1x_2}m_Bb_2) \nonumber \\ 
		&+ (b_1 \leftrightarrow b_2)], \nonumber \\
	h_{na}^2&(x_1,x_2,b_1,b_2)=K_0(\sqrt{x_1+x_2-x_1x_2}m_Bb_1) \nonumber \\ 
		&\times[\theta(b_1-b_2) K_0(-i\sqrt{x_1x_2}m_Bb_1)I_0(-i\sqrt{x_1x_2}m_Bb_2) \nonumber \\ 
		&+ (b_1 \leftrightarrow b_2)], 
\end{align}
}
where $K_0$ and $I_0$ are the Bessel functions. 

To suppress the higher-order corrections, the renormalization scale should be chosen as the maximum virtuality in each diagrams. 
The scales $t$ are shown below: 
{
\allowdisplaybreaks 
\begin{align}
	t_e^1=&\max(\sqrt{1-x_1}m_B, 1/b_0, 1/b_1), \nonumber \\
    t_e^2=&\max(\sqrt{x_0}m_B, 1/b_0, 1/b_1), \nonumber \\
	t_a^1=&\max(\sqrt{x_2}m_B, 1/b_1, 1/b_2), \nonumber \\
	t_a^2=&\max(\sqrt{1-x_1}m_B, 1/b_1, 1/b_2), \nonumber \\
	t_{ne}^1=&\max(\sqrt{x_0(1-x_1)}m_B, \sqrt{x_2(1-x_1)}m_B, \nonumber \\
		&\qquad 1/b_0, 1/b_2), \nonumber \\
    t_{ne}^2=&\max(\sqrt{x_0(1-x_1)}m_B, \sqrt{(1-x_2)(1-x_1)}m_B, \nonumber \\ 
		&\qquad 1/b_0, 1/b_2), \nonumber \\
	t_{na}^1=&\max(\sqrt{(1-x_1)x_2}m_B, 1/b_0, 1/b_1), \nonumber \\
    t_{na}^2=&\max(\sqrt{(1-x_1)x_2}m_B, \sqrt{1-x_1+x_1x_2}m_B, \nonumber \\
		&\qquad 1/b_0, 1/b_1). 
\end{align}
}

Conventionally, the threshold resummation factor $S_t(x)$ is parameterized to the following form \cite{Kurimoto:2001zj}: 
\begin{align} \label{eq:st}
  S_t(x)=\frac{2^{1+2c}\Gamma(3/2+c)}{\sqrt{\pi}\Gamma(1+c)}[x(1-x)]^c,
\end{align}
with $c=0.3$.

The Sudakov factors for the $B$ meson and the light meson $M$ are expressed as 
\begin{align} \label{eq:sudakovBandM}
	S_B(t)=&s(x_0,b_0,m_B)-\frac{1}{\beta_1} \ln\frac{\ln(t/\Lambda_{\textup{QCD}})}{\ln(1/(b_0\Lambda_{\textup{QCD}}))}, \nonumber \\
	S_M(t)=&s(x_1,b_1,m_B)+s(1-x_1,b_1,m_B) \nonumber \\
		&-\frac{1}{\beta_1}\ln\frac{\ln(t/\Lambda_{\textup{QCD}})}{\ln(1/(b_1\Lambda_{\textup{QCD}}))}, 
\end{align}
Up to the next-to-leading order, the factor $s(x,b,Q)$ derived from the resummation of double logarithms is defined as follows \cite{Li:1992nu, Li:1994zm}: 
{
\allowdisplaybreaks 
\begin{align} \label{eq:sudakov}
    &s(x,b,Q) \nonumber \\ 
		&=\frac{A^{(1)}}{2\beta_1}\hat{q}\ln\left(\frac{\hat{q}}{\hat{b}}\right)-\frac{A^{(1)}}{2\beta_1}\left(\hat{q}-\hat{b}\right)
			+\frac{A^{(2)}}{4\beta_1^2}\left(\frac{\hat{q}}{\hat{b}}-1\right)  \nonumber \\
		&\quad -\left[\frac{A^{(2)}}{4\beta_1^2}-\frac{A^{(1)}}{4\beta_1}\ln\left(\frac{e^{2\gamma_E-1}}{2}\right)\right] \ln\left(\frac{\hat{q}}{\hat{b}}\right) \nonumber \\
		&\quad +\frac{A^{(1)}\beta_2}{4\beta_1^3}\hat{q}\left[\frac{\ln(2\hat{q})+1}{\hat{q}}-\frac{\ln(2\hat{b})+1}{\hat{b}}\right] \nonumber \\
		&\quad +\frac{A^{(1)}\beta_2}{8\beta_1^3}\left[\ln^2(2\hat{q})-\ln^2(2\hat{b})\right] \nonumber \\
		&\quad +\frac{A^{(1)}\beta_2}{8\beta_1^3}\ln\left(\frac{e^{2\gamma_E-1}}{2}\right)\left[\frac{\ln(2\hat{q})+1}{\hat{q}}-\frac{\ln(2\hat{b})+1}{\hat{b}}\right] \nonumber \\
		&\quad -\frac{A^{(1)}\beta_2}{16\beta_1^4}\left[\frac{2\ln(2\hat{q})+3}{\hat{q}}-\frac{2\ln(2\hat{b})+3}{\hat{b}}\right] \nonumber \\
		&\quad -\frac{A^{(1)}\beta_2}{16\beta_1^4}\frac{\hat{q}-\hat{b}}{\hat{b}^2}\left[2\ln(2\hat{b})+1\right] \nonumber \\
		&\quad +\frac{A^{(2)}\beta_2^2}{1728\beta_1^6}\left[\frac{18\ln^2(2\hat{q})+30\ln(2\hat{q})+19}{\hat{q}^2} \right. \nonumber \\
		&\qquad -\left.\frac{18\ln^2(2\hat{b})+30\ln(2\hat{b})+19}{\hat{b}^2}\right] \nonumber \\
		&\quad +\frac{A^{(2)}\beta_2^2}{432\beta_1^6}\frac{\hat{q}-\hat{b}}{\hat{b}^3}\left[9\ln^2(2\hat{b})+6\ln(2\hat{b})+2\right],
\end{align}
}
where $\gamma_E$ is the Euler constant and $\Lambda_{\text{QCD}}$ is the QCD scale parameter. 
For the simplicity of the expressions, the variables $\hat{q}$ and $\hat{p}$ are defined by 
\begin{align} \label{eq:qphat}
  \hat{q}\equiv \ln\left(xQ/\left(\sqrt{2}\Lambda_{\textup{QCD}}\right)\right),\quad
  \hat{b}\equiv \ln\left(1/\left(b\Lambda_{\textup{QCD}}\right)\right).
\end{align}
The coefficients $\beta$ and $A$ are 
\begin{align} \label{eq:betaandA}
	\beta_1 &= \frac{33-2n_f}{12}, \quad \beta_2=\frac{153-19n_f}{24}, \quad A^{(1)} = \frac{4}{3}, \nonumber \\ 
	A^{(2)} &= \frac{67}{9}-\frac{\pi^2}{3} -\frac{10}{27}n_f+\frac{8}{3}\beta_1\ln\left(\frac{e^{\gamma_E}}{2}\right), 
\end{align}
with the number of activated flavors $n_f$.

\end{document}